\documentclass[journal,transmag]{IEEEtran}
\usepackage[top=0.62in, bottom=0.62in, left=0.62in, right=0.62in]{geometry}
\usepackage{amsmath}
\usepackage{multirow}
\usepackage{color}
\usepackage{float}
\usepackage{cite}
\usepackage[utf8]{inputenc}
\usepackage{lmodern}
\usepackage{tabularx}
\usepackage[bottom]{footmisc}
\usepackage{blindtext}
\usepackage{wrapfig}
\usepackage{graphicx}
\DeclareGraphicsExtensions{.eps,.pdf,.gif,.png,.jpg}
\usepackage{amssymb}
\usepackage{amsthm}
\usepackage{array}\usepackage{algorithm,algorithmic}
\usepackage{booktabs}
\usepackage{graphicx}
\usepackage{epstopdf}\usepackage[utf8]{inputenc}
\usepackage[hyphens,spaces,obeyspaces]{url}
\usepackage[english]{babel}
\usepackage{verbatim} 
\usepackage{lineno,nohyperref}
\usepackage{csquotes}
\usepackage{epstopdf} 
\modulolinenumbers[5]

\usepackage{ragged2e}
\usepackage{url} 
\usepackage[T1]{fontenc}
\usepackage[utf8]{inputenc}
\usepackage{microtype} 

\newcommand*{\affaddr}[1]{#1} 
\newcommand*{\affmark}[1][*]{\textsuperscript{#1}}
\newcommand{\vect}[1]{\boldsymbol{#1}}
\newcommand*{\email}[1]{#1}
\PassOptionsToPackage{hyphens}{url}
\newtheoremstyle{mystyle}
{}
{}
{\itshape}
{}
{\bfseries}
{.}
{ }
{\thmname{#1}\thmnumber{ #2}\thmnote{ (#3)}}

\theoremstyle{mystyle}
\newtheorem{remark}{Remark}

\newcounter{subassumption}[asu]

\makeatletter
\renewcommand{\p@subassumption}{\theasu}
\makeatother
\usepackage{amsthm}
\usepackage{xpatch,amsthm}
\makeatletter
\xpatchcmd{\@thm}{\fontseries\mddefault\upshape}{}{}{} 
\makeatother

\makeatother
\usepackage{cite}
\usepackage{amsmath,amssymb,amsfonts}
\usepackage{algorithmic}
\usepackage{graphicx}
\usepackage{textcomp}
\usepackage{xcolor}
\def\BibTeX{{\rm B\kern-.05em{\sc i\kern-.025em b}\kern-.08em
		T\kern-.1667em\lower.7ex\hbox{E}\kern-.125emX}}

\begin{document}

\title{Energy-Efficient Multi-Radio Microwave and IAB-Based Fixed Wireless Access for Rural Areas\\
		\thanks{This work was supported by NSERC (under project ALLRP 566589-21) and  InnovÉÉ (INNOV-R program) through the partnership with Ericsson and ECCC.
		We thank the Ericsson's Montréal
GAIA team for their constructive and helpful comments, which have
significantly improved the quality and clarity of this manuscript.}}
\author{%
Anselme Ndikumana\affmark[1], Kim Khoa Nguyen \affmark[1], Adel Larabi\affmark[2], and Mohamed Cheriet\affmark[1]\\
\affaddr{\affmark[1]Synchromedia Lab, École de Technologie Supérieure, Université du Québec, QC, Canada\\  \email{\{anselme.ndikumana, kim-khoa.nguyen; Mohamed.Cheriet\}}@etsmtl.ca}\\
\affaddr{\affmark[2] GAIA  Montreal, Ericsson Canada}\\
\email{\{adel.larabi\}@ericsson.com}\\
}

\maketitle

\begin{abstract}
\label{sec:abstract} 
Deploying fiber optics as a last-mile solution in rural areas is not economically viable due to low population density. Nevertheless, providing high-speed internet access in these regions is essential to promote digital inclusion. 5G Fixed Wireless Access (5G FWA) has emerged as a promising alternative; however, its one-hop topology limits coverage. To overcome this limitation, a multi-hop architecture is required. This work proposes a unified multi-hop framework that integrates long-haul microwave, Integrated Access and Backhaul (IAB), and FWA to provide wide coverage and high capacity in rural areas. As the number of hops increases, total energy consumption also rises, a challenge often overlooked in existing literature. To address this, we propose an energy-efficient multi-radio microwave and IAB-based FWA framework for rural area connectivity. When the network is underutilized, the proposed approach dynamically operates at reduced capacity to minimize energy consumption. We optimize the  off, start-up, serving, deep sleep, and wake-up sates of microwave radios to balance energy use and satisfying data rate requirements. Additionally, we optimize resource block allocation for IAB-based FWA nodes connected to microwave backhaul. The formulated optimization problems aim to minimize the energy consumption of long-haul microwave and multi-hop IAB-based network while satisfying data rate constraints. These problems are solved using dual decomposition and multi-convex programming, supported by dynamic programming. Simulation results demonstrates that the proposed approach significantly reduces energy consumption while satisfying required data rates. 
\end{abstract}

\section{Introduction}
\label{sec:introduction} 
\subsection{Background and Motivations}
\label{subsec:motivation} 
Connectivity to high-speed internet, especially to 5G networks, is not equal to all populations. Some parts of the world remain unconnected or connected to lower-speed internet such as 3G, especially in rural areas. According to the International Telecommunication Union (ITU),  $33$ \% of the global population was unconnected in 2023 \cite{ITU}. Improving high-speed internet access in rural areas is crucial for economic development, education, and quality of life. Laying fiber optic cables in rural areas can provide high-speed internet with excellent reliability and bandwidth, but it is costly due to the lower population density and there is no guarantee of return on investment. 5G FWA is an affordable
solution to provide Internet access to these areas. In 5G FWA, the houses are equipped with Customer Premises Equipment (CPE)  with an antennas mounted on the roofs of the houses that are wirelessly connected to fixed cellular base stations \cite{Maravedis}. FWA is growing solution, where it is forecasted to be $300$ million FWA connections by 2028 \cite{ericsson2024Mobility}.

Millimeter-wave (mmWave) can enhance the capacity of  5G FWA  \cite{chaudhuri2021extended}. However, mmWave signals suffer from high propagation loss and limited penetration capabilities, resulting in a smaller coverage area. When used in 5G FWA, which typically operates as a one-hop solution, this limitation restricts the overall coverage. Consequently, additional technologies are needed to extend the one-hop architecture of 5G FWA into a multi-hop wireless solution. The authors in \cite{hashemi2017integrated} combined Integrated Access and Backhaul (IAB) and FWA as a unified solution to cover large areas. The IAB  network \cite{3GPP38874} comprises an IAB donor and IAB nodes. The IAB donor is the base station connected to the core network via a fiber optic connection. The other base stations in the IAB network are IAB nodes, which connect to the IAB donor through wireless backhaul. In the context of IAB-based FWA, CPE can connect to either the IAB donor or the IAB nodes using wireless links. Each IAB node includes a Distributed Unit (DU) and a Mobile Termination (MT) unit, which serves downstream CPEs and  IAB-MTs. Furthermore, the IAB-MT allows an IAB node to function like a relay node when connected to its parent IAB-DU. The IAB donor has both a DU and a Control Unit (CU). 

Instead of relying on fiber optic, a microwave node can connect the IAB donor to the core network. Long-haul microwave links can deliver up to 10 Gbps per hop \cite{ericsson2024}. When Long-haul microwave combined with IAB and FWA in a unified multi-hop architecture that uses high-frequency bands, the network can cover large areas and deliver high capacity to rural areas. Long-haul microwave radios operating in the 11 GHz and 71–86 GHz to 191.7–194.8 THz ranges have been proposed to provide broadband access to agricultural farms \cite{zu2023arahaul}, achieving up to 160 Gbps over distances of up to 16 km. However, such an integrated approach combining long-haul microwave, IAB, FWA, and high-frequency bands remains largely unexplored in the literature.
\subsection{Challenges in Providing Long-haul Microwave
and  IAB-based FWA Network to Rural Areas}
\label{subsec:challenges} 
The key challenges faced by long-haul microwave and multi-hop IAB-based FWA network in rural areas are:
\begin{itemize}
	\item Integrated long-haul microwave and IAB-based FWA as unified network can extend   network coverage in rural areas. However, energy consumption increases with the number of hops, and existing studies \cite{zhang2021resource, yu2023coordinated, zu2023arahaul, begishev2021performance} have not addressed this energy consumption issue.
	\item In multi-radio and multi-band microwave backhaul for IAB-based FWA, it is challenging to decide when and which radios to place in deep sleep  for saving energy during low network utilization, and when to reactivate radio to meet capacity demands. Improper timing can degrade network service quality, especially during sudden network traffic surges \cite{frithiofson2022energy}.
	\item Rural areas have lower population density compared to urban and suburban regions, making it likely that radio resources allocated to CPEs/MTs in IAB-based FWA are often underutilized \cite{ndikumana2024renewable}.
\end{itemize}
\subsection{Contributions}
\label{subsec:contribution}
We propose an energy-efficient multi-radio microwave and IAB-based FWA framework for rural areas to address the challenges outlined above. Our main contributions are as follows:
\begin{itemize}
	\item We consider multiple radios at each microwave node. When microwave backhaul is underutilized, one or more radios can switch to energy-saving states. Unlike existing studies that only consider off, on, and deep sleep states, we introduce five operational states: completely off, startup, serving, deep sleep, and wake-up, which are novel in the microwave communication literature.
	\item We formulate the power management problem as an optimization model that determines the optimal state–action pairs for long-haul microwave to minimize  energy consumption while satisfying the data rate requirements of IAB-based FWA in rural areas.
	\item For IAB-based FWA connected to microwave backhaul, we consider both mmWave and mid-band frequencies to balance coverage and capacity in rural area deployment. By considering 5G mixed numerologies and maximum transmission bandwidths, we propose a dynamic resource block allocation scheme and formulate an optimization problem to minimize the energy consumption of multi-hop IAB-based FWA connected via long-haul microwave links.
	\item We join the formulated optimization problems for long-haul microwave and IAB-based FWA as an unified optimization problem that jointly satisfies downlink data rate requirements  while minimizing energy consumption. Then,  we use dual decomposition method and disciplined multi-convex programming (DMCP) \cite{shen2017disciplined, boyd2011distributed}, supported by dynamic programming \cite{van2012reinforcement} as solution approach.
\end{itemize}

The remainder of this paper is organized as follows. Section \ref{sec:literaturereview} reviews related work, and Section \ref{sec:system_model} presents the system model. Section \ref{sec:problem_formulation} formulates the problem, while Section \ref{sec:Proposed_Solution} describes the proposed solution. Section \ref{sec:simulation_results_analysis} provides the performance evaluation, and Section \ref{sec:conclution} concludes the paper.

\section{Literature Review}
\label{sec:literaturereview}
We classify the related work into three main categories: (i) FWA and resource allocation, (ii) IAB-based FWA and resource allocation, and (iii) microwave backhauling and radio-link bonding.

\emph{FWA and Resource Allocation:}
In \cite{rahmawati2022assessing}, the authors proposed a method to estimate the number of base stations required for FWA capacity and coverage planning, using an urban residential case study. In \cite{lappalainen2022planning}, the authors developed an approach to determine the maximum number of houses that can simultaneously connect to an FWA network while meeting minimum bit rate requirements, given network resources and cell radius, using MIMO-based fixed broadband service. In \cite{de2022outdoor}, mmWave channel modeling was applied for a $60$~GHz 5G FWA network. Similarly, \cite{de2023mmwave} analyzed FWA performance at $28$~GHz, $60$~GHz, and $140$~GHz, showing that higher frequencies suffer greater path loss but offer higher capacity due to larger bandwidths. In \cite{castellanos2023evaluating}, both urban and rural scenarios were evaluated for $60$~GHz FWA deployments, highlighting the role of high-frequency bands in achieving high capacity. Energy-aware FWA were considered in \cite{ndikumana2023digital, ndikumana2024renewable, ndikumana2024digital}, where base stations were powered by both the electrical grid and renewable energy sources, jointly optimizing energy consumption and radio resource allocation. However, conventional FWA remains a one-hop solution, and high-frequency bands are prone to significant path loss, limiting rural coverage unless more base stations are deployed to reduce the distance to CPEs.

\emph{IAB-based FWA and Resource Allocation:}
In \cite{hashemi2017integrated}, IAB was combined with $28$GHz FWA to deliver high capacity to residential areas. UAVs as IAB nodes were proposed in \cite{tafintsev2023airborne} to enhance flexibility in topology configuration. To address mmWave path loss, \cite{begishev2021performance} proposed a multi-band microwave–mmWave solution, rerouting traffic to sub-6GHz links when mmWave links fail. Similarly, \cite{yao2022delay} explored integrated mmWave and sub-6~GHz systems to overcome blockage and intermittent connectivity, formulating a packet scheduling problem across both interfaces. The work in \cite{yu2023coordinated} compared centralized and distributed resource coordination in IAB, proposing a hybrid coordinated approach. While these designs improve connectivity, energy consumption and latency both increase with the number of hops. Importantly, none of these works address energy reduction in multi-hop IAB-based FWA, especially in rural deployments where radio resources allocated to CPEs/MTs are often underutilized.
\begin{figure*}[t]
	\centering	\includegraphics[width=1.8\columnwidth]{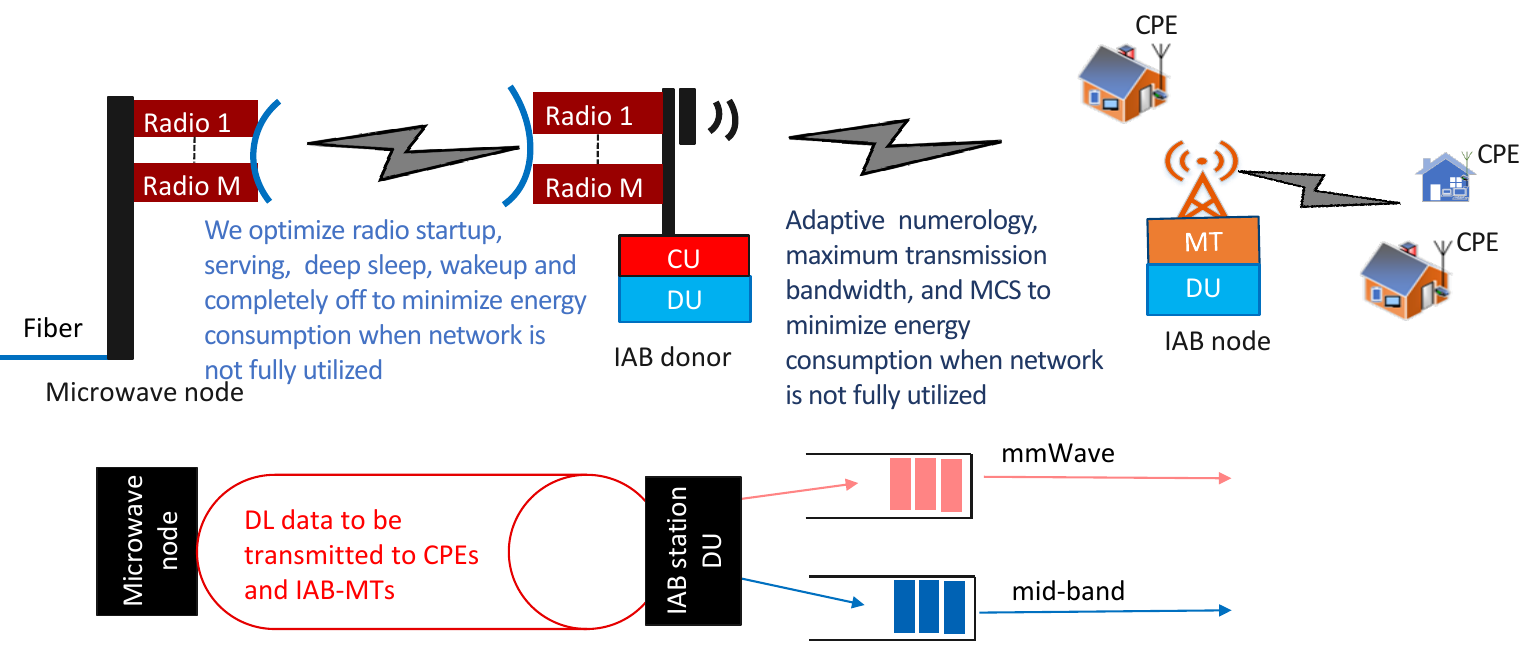}
	\caption{Illustration of our system model.}
	\label{fig:systemmodel2}
\end{figure*}

\emph{Microwave Backhauling and Radio-link Bonding:}
Microwave backhaul remains a critical transport solution above $6$~GHz, as noted in \cite{sellin2020enhancing}. Long-haul microwave radios operating in $11$~GHz, $71$–$86$~GHz, and up to $191.7$–$194.8$THz were evaluated in \cite{zu2023arahaul} for rural broadband access in agricultural areas. Radio-link bonding, aggregating capacity across multiple radios, was studied in \cite{wang2022triple} using triple-band scheduling (28GHz, E-band, and THz), and in \cite{colzani2022long} with high-power amplifier modules for long-reach E-band links. Energy saving in microwave backhaul was discussed in \cite{frithiofson2022energy}, which noted that utilization often stays below 50\%, suggesting opportunities to power down radios during low traffic. However, in multi-radio and multi-band settings, determining when to switch radios into deep sleep and when to wake them without service disruption remains a complex challenge.

From the above review, long-haul microwave and multi-hop IAB-based FWA for rural areas has not been jointly addressed in the literature. To the best of our knowledge, this is the first study to minimize the energy consumption of such networks, incorporating multi-band microwave links and 5G mixed numerologies.

\section{System Model}
\label{sec:system_model} 

\begin{table}[t]
	\caption{Summary of key notations.}
	\label{tab:table1}
	\begin{tabular}{ll}
		\toprule
		Notation & Definition\\
		\midrule
		$\mathcal{J}$ & Set of microwave nodes, $|\mathcal{J}|= J$\\
		$\mathcal{M}$ & Set of microwave radios  $|\mathcal{M}|= M$\\
		$\mathcal{V}$ & Set of CPEs and IAB-MTs, $|\mathcal{V}|= V$\\
		$\mathcal{N}$ & Set of IAB DUs, $|\mathcal{N}|= N$\\
        $d_v$ & Minimum data rate for CPE or IAB-MT $v$\\
		$D^j_m$ & Data rate for radio m of microwave node $j$ \\
            $D$ & Total data rate for IAB-based FWA \\
  $\vect{x}$ & Vector of decision variable $x^j_{m, k} $ \\
  $\mathcal{A}^j$ & Action space for microwave node $j$ \\
  $\Omega^j$ & State space for microwave node $j$ \\
  $\Phi(\Omega^j, \mathcal{A}^j)$ & Transition matrix from one state to another\\
 $t^j_{m,4}$ & Time for microwave radio to complete the startup\\
  $ t^j_{m,5}$ &  Time for microwave radio  to complete the wake-up\\
$ D_{\text{Max}}^{n, t}$ &  Maximum data rate that each IAB-DU $n$ \\
$y^n_{mm}$  &  Decision variable for selecting mmWave  \\
$y^n_{md}$  & Decision variable for selecting mid-band  \\
$\Psi_{rds}$ & DL traffic status for the radio to go into deep sleep  \\
		\bottomrule
	\end{tabular}
\end{table}

We present our system model in Fig \ref{fig:systemmodel2}, while the key notations are shown in Table \ref{tab:table1}.

Our system model considers the microwave link between the microwave node and the IAB donor, where the microwave node is connected to the Core Network (CN) using optical fiber. Each microwave node is equipped with one or more radios. We denote $\mathcal{M}$ as the set of microwave radios and $\mathcal{J}$ as a set of microwave nodes,  where $M^j$ is the number of radios associated with the microwave node $j \in \mathcal{J}$. We consider multi-bands, where microwave radios can use bands in narrow ($6-15$ $Ghz$), wide ($18-42$ $Ghz$), or very wide ($71-76$ $Ghz$ or $81 - 86$ $Ghz$) bands depends on demands.  The IAB donor has microwave radios for backhauling and other radio(s) for serving IAB nodes and CPEs for internet access. We use IAB station to mean IAB donor or IAB node, where each IAB station has duo bands for
mid-band and mmWave interfaces. Furthermore,  DU of IAB station  has a MAC scheduler that performs resource block (RB) scheduling, where we consider two buffers that store transient packets waiting to be served by either the mmWave interface or
the mid-band interface. In other words, we have two outgoing packet buffers for mmwave and midi-band. Since we focus on RB scheduling and MCS at DU, here, we denote $\mathcal{N}$ as a set of DUs. Since each IAB station has O-DU, Unless stated otherwise, we use the terms DU and IAB station interchangeably. Also, each DU can serve CPE and IAB-MT, we use the term terminal to mean CPE or IAB-MT. We denote $\mathcal{V}$ as a set of terminals, i.e., IAB-MT and CPEs. The relation between long-haul microwave,  IAB, and FWA is shown in Fig. \ref{fig:systemmodel2}.
\begin{figure}[t]
	\centering	\includegraphics[width=1.0\columnwidth]{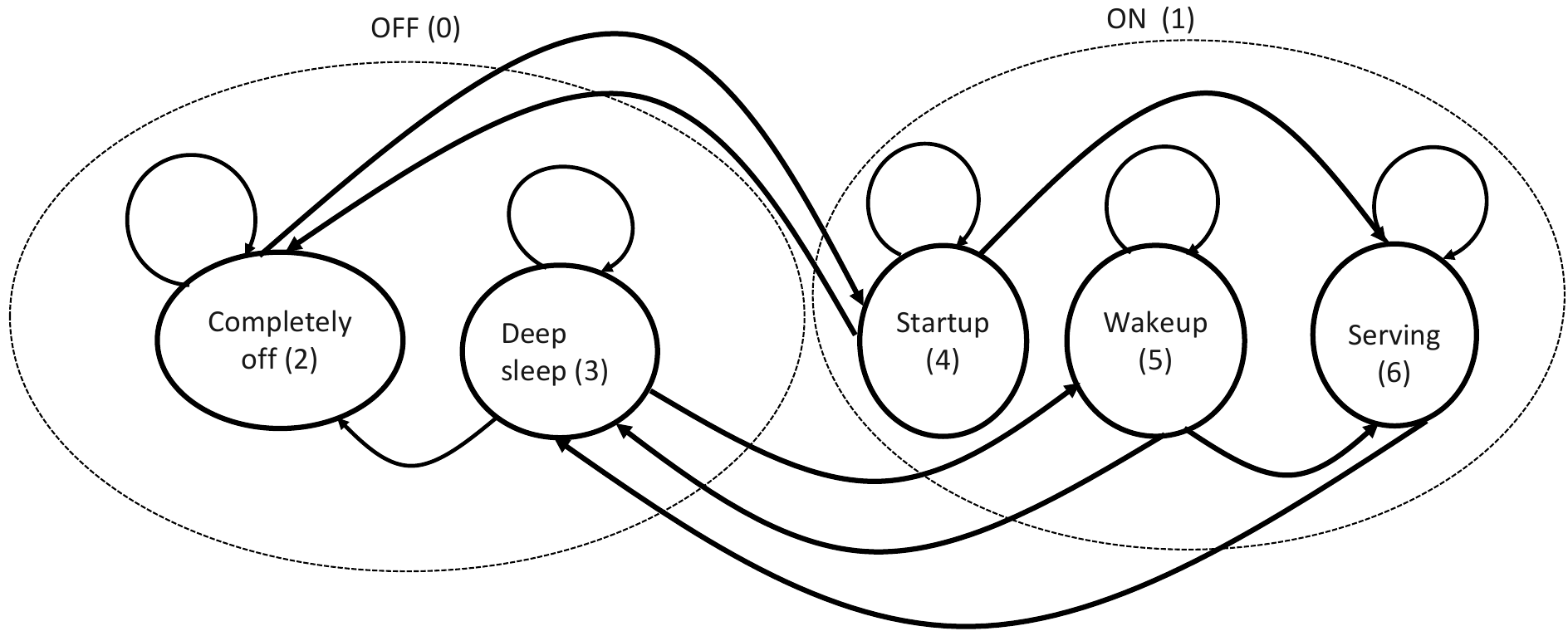}
	\caption{Microwave radio states and sub-states.}
	\label{fig:mdp_Model}
\end{figure}

\subsection{Multi-radio Microwave Backhaul for 5G IAB-based FWA Network}
\label{subsec:MicrowaveBackhualing} 
As shown in Fig. \ref{fig:systemmodel2}, the DL data reaches the IAB-based FWA network via microwave backhauling. We denote $d_v$ as the data rate that needs to be sent to CPE or IAB-MT $v$.  Since we have multiple terminals, the DL data can be denoted as $D$, where $D =\sum_{v=1}^{V} d_v$. The data $D$ passes through nodes with multiple  microwave radios, i.e., 
 multiple microwave links to reach IAB donor. 

For a long-haul microwave, we consider two physical states for microwave radio, where microwave radio can be in OFF(0) or  ON (1) state. Furthermore, we consider the microwave node to have a controller.  At the controller, we consider 5 sub-states: completely off, radio deep sleep, startup, and serving. From a controller's point of view, in a deep sleep sub-state, the controller sends a command to the microwave radio to go OFF.  In the startup sub-state, the controller sends a command to the microwave radio to go ON. 

 In Fig. \ref{fig:mdp_Model}, OFF (0) state, the microwave controller considers deep sleep (2) and completely off (3) sub-states, where the DC power is switched off. However,
to prevent moisture buildup in a microwave radio in deep sleep configuration, it must be powered for a certain period every day. Thus, we distinguish between deep sleep and completely off. Completely off refers to when the unit is either new or has been unused in a serving state for an extended period, such as seven consecutive days. A microwave radio in a completely off state can be relocated to another area. In completely off, the microwave radio consumes the power $P^j_{{m,2}}(t^j_m)=0$. In deep sleep, the microwave radio consumes the power  $P^j_{{m,3}}(t^j_m)$.

In the On (1) state, the controller considers startup (4), wake-up (5), and serving (6) sub-states. We use $P^j_{{m,4}}(t^j_m)$ to denote the power required to startup, $P^j_{{m,5}}(t^j_m)$ as the power required for wake-up, and $P^j_{{m,6}}(t^j_m)$ as the power required in the serving sub-state.

Based on Fig. 2, in combining microwave radio state and controller sub-states, we use $K$ as the total number of states.
The state $\{ 0\{2, 3\}$  and $1\{4,5, 6 \}\} \times \{1,2, \dots, M^j\}$ can be  generalized as 
$\Omega^j=\{\Omega^j_{m, k}\}| m=1, 2, \dots, M^j$ and $k=0, 1, 2, \dots, K$.
We define $\vect{x}=\{x^j_{m, k} \}$ as a vector of decision variables that indicate the states microwave radio belongs to.

 
 Let us consider $m$ and $n$ as two microwave radios in serving sub-states. The achievable SNR $\delta_{m}$  can be expressed as follows:
\begin{equation}
	\label{eq:AchSNR}
	\delta_{m} = 
	\frac{|G^{RX}_{m,6} |^2  P^{TX}_{n,6} }{\sigma_{mn}^2},\; 
\end{equation}
where $P^{TX}_{n,6}$ is the allocated transmission power, $G^{RX}_{m,6}$ is the channel gain, and $\sigma_{mn}^2$ represents the noise power. Based on the achievable SNR, the maximum achievable DL data rate for radio $m$ for microwave node $j$ can be expressed as:
\begin{equation}
	\label{eq:data_rate}
	\begin{aligned}
		D^j_m=\omega_m^{b}\log_2\left(1 + \delta_{m} \right), 
	\end{aligned}
\end{equation}
where $\omega_m^{b}$ is the channel bandwidth. Considering all microwave radios $M^j$ used at microwave node $j$. The data rate should satisfy the following constraint:
\begin{equation}
\label{eq:data_rate_satisfaction}
\sum_{m=1}^{M^j} D^j_m \geq D.
\end{equation}
In the multi-radios environment, when $D$ is not high, we can reduce the number of microwave radios needed to satisfy (\ref{eq:data_rate_satisfaction}).

At the time $t^j_m$, when the existing microwave radio(s) in the serving sub-state can not satisfy (\ref{eq:data_rate_satisfaction}), the controller can consider switching the existing microwave radio(s) in the completely off to startup sub-states. In other words, switching from completely off to startup sub-states happens when:
\begin{equation}
\sum_{m=1}^{{M}^{j}} x^j_{m, 6}D^j_m (t^j_m) < D (t^j_m), D (t^j_m) > 0.
\end{equation}
Therefore, we define an action $a^{1,j}_{m,4}$ of switching from completely off to startup sub-state. Otherwise, action $\tilde{a}^{0,j}_{m,2}$ is taken, where microwave radio remains in a completely off sub-state. By taking action $a^{1,j}_{m,4}$, the controller sends a command for the microwave radio to switch from OFF state to ON state. This action is denoted $a^{1,j}_{m}$.

When the microwave radio starts, it stays in the startup state for:
\begin{equation}
\sum_{t^j_m=1}^{\tau}x^j_{m, 4}t^j_m\leq t^j_{m,4},
\end{equation}
where $t^j_{m,4}$ is the time required so that the microwave radio completes the startup.
 When the microwave radio $m$ is in a start-up sub-state  and cannot complete the start-up process in time $t^j_{{m},4}$, we consider that it fails and needs to return to the completely off sub-state. Then, this action is denoted $\tilde{a*}^{0,j}_{{m},2}$. Otherwise, we denote $a^{1,j}_{m,6}$ as the action of switching from start-up sub-state to serving sub-state for microwave radio in ON state. 

When network traffic is very low, unused microwave radio in a serving sub-state can go to deep sleep. Therefore, we define data traffic $\Tilde{D}_{rds}$ as radio deep sleep threshold and $\Psi_{rds}$ as DL traffic status for the radio to go into deep sleep sub-states. The $\Psi_{rds}$ is given by:
\begin{equation}
\Psi_{rds} =\max \{(\sum_{m=1}^{M^j}x^j_{m, 6} D^j_m (t^j_m) -D (t^j_m)), \Tilde{D}_{rds} \}.
\end{equation}
When $\Psi_{rds} = \Tilde{D}_{rds}$, we consider switching from the serving sub-state to the deep-sleep sub-state for microwave radio in the ON state. We denote this action as $a^{0,j}_{m,3}$. By taking action $a^{0,j}_{m,3}$, the controller also sends a command for the microwave radio to switch to OFF state. This action is denoted $a^{O,j}_m$.   Otherwise, microwave radio remains in the serving sub-state by taking action $\tilde{a}^{1,j}_{m,6}$.

When the network traffic increases again, the microwave in deep sleep sub-state can wake up. Therefore, we define data traffic $\Tilde{D}_{w}$ as the radio wake-up threshold and $\Psi_{w}$ as DL traffic status for wake-up. DL traffic status is formulated as follows:
\begin{equation}
\Psi_{w} =\max \{(\sum_{m=1}^{M^j}x^j_{m, 3} D^j_m (t^j_m) -D (t^j_m)), \Tilde{D}_{w} \}.
\end{equation}
 When $\Psi_{w} = \Tilde{D}_{w}$, we consider switching from deep sleep sub-state to the wake-up sub-state. We denote this action as $a^{1,j}_{m,5}$. Otherwise, microwave radio remains deep sleep sub-state by taking action $\tilde{a}^{0,j}_{m,3}$. We consider $ t^j_{m,5}$ as the time required so that the microwave radio completes the wake-up such that:
 \begin{equation}
 	\sum_{t=1}^{\tau}\vect{x}^j_{m, 5}t^j_m\leq t^j_{m,5}.
 \end{equation}
When the microwave radio $m$ is in a wakeup sub-state and cannot complete the wakeup process in time $t^j_{{m},5}$, we consider that it fails and needs to return to the deep sleep sub-state. This action is denoted $\tilde{a*}^{0,j}_{{m},3}$. Otherwise, we consider switching from wake-up sub-state to the serving sub-state for microwave radio in the ON state. This action is denoted $a^{1,j}_{m,7}$. 

When the microwave radio is not used for a long period $T$, it can be switched to the completely off sub-state. In other words, the completely off sub-sate happens when the following condition is satisfied:
\begin{equation}
\sum_{t=1}^{T} x^j_{m,3} D^j_m (t^j_m)=0.
\end{equation}

We denote $a^{0,j}_{m,2}$ as an action of switching from deep sleep to the completely off sub-state when $\sum_{t=1}^{T} x^j_{m,3} D^j_m (t^j_m)=0$. Otherwise, this action $\tilde{a}^{0,j}_{m,3}$ is taken, where the microwave radio remains deep sleep sub-state.

Based on the actions defined above, we define $\mathcal{A}^j=\{a^{s,j}_{m, k},  a^{1,j}_{m},a^{0,j}_{m}\}| m=1, 2, \dots, M^j$, $s=0,1 $, and $k=0, 1, 2, \dots, 7$ as an action space. Furthermore, we define $\Phi(\Omega^j, \mathcal{A}^j)$ as a transition matrix from one state to another.

The neighboring microwave nodes need to synchronize the states of their radios. Also, each microwave node should always have at least one radio in the serving state. The upstream controller of upstream microwave node sends its radio states  to the downstream controller of the downstream microwave node. Then, the downstream  microwave controller uses its radio states and the received upstream radio states to match the states of radios. As an example, if the radio 1 of the upstream microwave node is in a deep sleep sub-state, the radio 1 of the downstream  microwave node should be in a deep sleep sub-state as well.

\subsection{IAB-based FWA Network Connected to Multi-radio Microwave Backhaul}
\label{subsec:FAWBackhualing} 

\begin{figure}[t]
	\centering	\includegraphics[width=1.0\columnwidth]{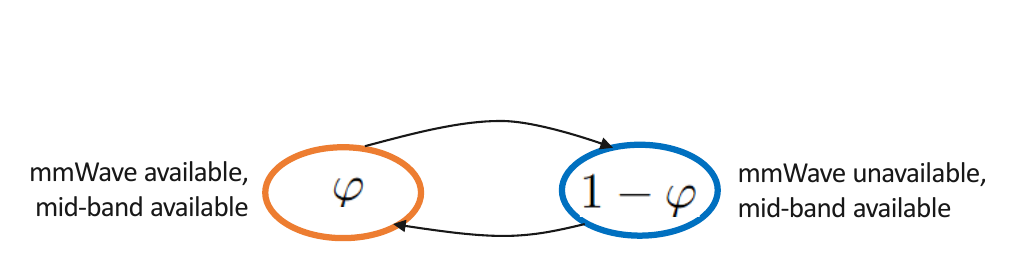}
	\caption{Mid-band and mmWave states at IAB station.}
	\label{fig:Two_statesIAB}
\end{figure}

Once the data $D$ reaches the IAB donor, it can use mid-band and microwave to send the data to its children nodes such as CPEs and IAB-MTs.
As shown in Fig. \ref{fig:Two_statesIAB}, we model the mid-band and microwave of the IAB-node station as a two-state  system.The first state represents both mmWave and mid-band being available. The second state represents mmWave being unavailable while mid-band remains available. We ignore the state where both mmWave and mid-band are unavailable. Furthermore, due to the location of the IAB station, there may be areas where the mid-band is available, but mmWave is not.
\begin{equation}
O_n=
\begin{bmatrix}
\varphi & 1-\varphi\\
\end{bmatrix},
\end{equation}
where $\varphi$  denotes  the state:  mmWave is available and
mid-band available. Otherwise, $1- \varphi$ denotes the state: mmWave unavailable and 
mid-band available. We defined the decision variable $y^n_{mm}$ for selecting mmWave over mid-band as follows:
\begin{equation}
	\setlength{\jot}{10pt}
	y^n_{mm}=
	\begin{cases}
		1,\; \text {if mmWave and mid-band}\\ \;\;\;\;\text {are available, i.e., in state $\varphi$}, \\
		0,\;\text{otherwise.}
	\end{cases}
\end{equation}
Furthermore, we define mid-band selection variable when mmWave is not available as follows:
\begin{equation}
	\setlength{\jot}{10pt}
	y^n_{md} =
	\begin{cases}
		1,\; \text {if mmwave is not available and}\\ \;\;\;\;\text {mid-band is available, i.e., in state $1-\varphi$}, \\
		0,\;\text{otherwise.}
	\end{cases}
\end{equation}
The IAB station should use mmWave or mid-band to connect its IAB-MT or CPE by satisfying the following constraint:
\begin{equation}
	\setlength{\jot}{10pt}
y^n_{md}+y^n_{mm}=1.
\end{equation}
Here, we assume that each IAB-MT and CPE reports Reference Signal Received Power (RSRP) and Reference Signal Received Quality (RSRQ) to its parent node. This helps the parent node to know the availability of mmWave and mid-band signals.

Based on the selected frequency band (mmWave or mid-band), the IAB station needs to allocate RB to each IAB-MT or CPE connected to it. We assume that the number of RBs is limited. Here, we remind that the number of RBs varies depending on the numerology $\mu$ and the maximum transmission bandwidth $\omega$. Therefore, we define $\beta^{u, \omega}$ as the total number of RBs of the selected band at the IAB station with numerology $\mu \in \mathcal{U}$ and transmission bandwidth $\omega \in \vect{\Omega}$.  We use $\mathcal{U}$ as a set of 5G numerology and $\vect{\Omega}$ as the set of maximum transmission bandwidth. For example, when transmission bandwidth $\omega=400$ MHz selected for mmWave and numerology $\mu=3$, we can have a maximum of $264$ RBs \cite{etsi138}.  We define  numerology selection decision variable $z^ {\omega}_{u}=1$ if numerology $\mu$ of bandwidth $\omega$ is selected. Otherwise, $z^ {\omega}_{u}=0$. At the time $t$, we  estimate the maximum data rate that each IAB-DU $n$ can handle using the following formulation defined in \cite{etsi1308}:
\begin{align}
     D_{\text{Max}}^{n, t} = 10^{-6}\sum_{j=1}^{J}	( \,\nu^{(j)}_{\text{layers}}Q^{(j)}_{mcs}  l^{(j)}  R_{max}  \frac{12 w^{n}_{\beta} \beta^{u, \omega}}{\mathcal{T}^{u} } \times\nonumber\\
	(1-{OH}^{(j)})). 
\end{align}
where $J$ represents the number of aggregated component carriers, and $\nu^{(j)}_{\text{layers}}$ is the maximum number of multiple-input multiple-output layers. $Q^{(j)}_{mcs}$ is the modulation order, and $l^{(j)}$ is a scaling factor.  ${OH}^{(j)}$ denotes the control channel overhead. Furthermore, $\mathcal{T}^{u}$ is the average orthogonal frequency division multiplexing (OFDM) symbol duration in a given subframe of selected numerology $\mu$. We denote $R_{max}$ as the maximum coding rate.

Using data $d_{v}$ from longhaul microwave backhauling that  IAB-DU $n$ needs to send to its CPE or an IAB-MT $v$, we define the number of RBs $ \beta_{v}^{{n}} (d_{v})$ required to deliver $d_{v}$ as follows: 
\begin{align}
	\label{Rb_2}
	\beta_{v}^{{n}} (d_{v})= \lceil
	\frac{10^{6} d_{v}  \mathcal{T}^{u}}{ \sum_{j=1}^{J}(\nu^{(j)}_{layers}  Q^{(j)}_{mcs}   l^{(j)} 12  R_{max}(1-{OH}^{(j)}))}\rceil.
\end{align}
In other words, $\beta_{v}^{{n}} (d_{v}) $ is RBs allocated to terminal $v$ at IAB-DU $m$ such that:
\begin{align}
	\sum_{v=1}^{|\mathcal{V}^n|}w^{n}_{v}	\beta_{v}^{{n}} (d_{v})  \leq \beta^{u, \omega} ,\;  \forall n  \in \mathcal{N}.
 \end{align}
We define $\mathcal{V}^n$ as the set of  terminals ( i.e., CPEs and IAB-MTs) connected to IAB-DU $n$.
We define RB allocation decision variable $w^{n}_{v}=1$ if RB $\beta_{v}^{{n}} (d_{v})$ is allocated to  terminal $v$. Otherwise, $w^{n}_{v}=0$.

We consider RB $\beta_{v}^{{n}} (d_{v})$ to be partitioned into $f$ in the frequency domain and $t$  in the time domain. The minimum time-domain length in RB can be the time for transmitting one OFDM symbol.  However, the exact time domain length varies depending on the Start and Length Indicator (SLIV) for the time domain related to the physical downlink shared channel.

The achievable SNR $\delta^v_{t,f}$ of each terminal $v$ on the allocated RB $\beta_{v}^{{n}} (d_{v})$ is given by:
\begin{equation}
	\label{eq:AchSNR}
	\delta^v_{t,f} = 
	\frac{|h^{v}_{t,f}|^2  p^v_{t,f} }{\sigma_v^2 },\; 
\end{equation}
where $p^v_{t,f}$ is the allocated transmission power of each RB $\beta_{v}^{{n}} (d_{v})$,  $h^{v}_{t,f}$ is the channel gain; and $\sigma_v^2$ represents the noise power. Based on the achievable SNR,  we define the following maximum achievable data rate for terminal $v$:
\begin{equation}
	\label{eq:data_rate}
	\begin{aligned}
		D_v^n =\omega^{v,\beta}\log_2\left(1 + \delta^v_{t,f}\right),  \;\forall v \in \mathcal{V},
	\end{aligned}
\end{equation}
where $\omega^{v,\beta}$ is the bandwidth of RB $\beta_{v}^{{n}} (d_{v})$.

IAB-DU should choose the MCS that matches the SNR and the wireless capacity to cope with the channel variations. A higher SNR allows a high-order modulation scheme to achieve higher data rate. A lower SNR requires a lower-order modulation scheme and allows lower coding rate. Therefore, each IAB-DU must select the MCS that satisfies the minimum bitrate $d_{v}$ while minimizing power consumption. In our approach, IAB-DU selects MCS index $i$ and minimum SNR threshold $\delta^{i, n}_{v,t}$ that satisfy $d_{v}$ as target SNR. Then, we define the MCS index selection variable $\kappa_{v}^{n}=1$ if MCS index $i$ is selected at IAB-DU $m$ that serves the terminal $v$. Otherwise, $\kappa_{v}^{n}=0$.

Based on SNR $\delta^v_{t,f}$, $\delta^{i, n}_{v,t}$, $ D_v^n$, and $d_v$, if $\delta^v_{t,f} >\delta^{i, n}_{v,t}$ and $ D_v^n \geq d_v $, IAB-DU $m$ decreases the transmit power such that $\delta^v_{t,f} =\delta^{i, n}_{v,t}$. if $\delta^v_{t,f} <\delta^{i, n}_{v,t}$ and $ D_v^n < d_v$, IAB-DU $n$ selects new MCS index $i$ and increases transmission power such that $D_v^n \geq d_v$. Furthermore, when $\delta^v_{t,f} =\delta^{i, n}_{v,t}$ and $D_v^n \geq d_v $, IAB-DU $n$ keeps MCS index $i$ and transmission power unchanged. 	The updated transmission power can be expressed as \begin{equation}
				p^v_{{t+1},f}=\frac{\delta^{i, n}_{v,t} w^{n}_{v}\kappa_{v}^{n}p^v_{t,f}}{ \delta^v_{t,f}}.
\end{equation}
Here, we assume that existing power control approaches, such as closed-loop power control, can be used to determine initial $p^v_{t,f} $. Furthermore, we define, at each IAB station $n$, the sum of transmit powers for all the scheduled RBs as \begin{equation}
p^n=\sum_{v=1}^{V^n}	(y^n_{md}+y^n_{mm}) p^v_{{t+1},f}.
\end{equation}

To track traffic load and RB utilization, we define the following RB utilization ratio $\Gamma_{t}^n$ at each IAB-DU $m$:
\begin{equation}
 \Gamma_{t}^n=\frac{\sum_{v=1}^{V^n}(y^n_{md}+y^n_{mm})w^{n}_{v} \beta_{v}^{{n}} (d_{v})}{ \beta^{u, \omega}  }.
\end{equation} 
Furthermore, each IAB-DU $m$ keeps tracking its queuing, i.e.,   arrival rate $\iota^n_{mm}$ for mmWave and $\iota^n_{md}$ for mind-band. It also tracks  service rate $g^n_{mm}$ for mmWave and $g^n_{md}$ for mind-band in DL. The  DL traffic load at each IAB-DU $m$ can be expressed as:
\begin{equation} 
	\rho^m_t= \frac{\sum_{v=1}^{V^n}(y^n_{md}\iota ^n_{md}+y^n_{mm}\iota^n_{mm})}{\sum_{v=1}^{V^n}(y^n_{md}g^n_{md}+y^n_{mm}g^n_{mm})}. 
\end{equation}	

The IAB station uses RB utilization to update the RB $\beta^{u, \omega}$, which imvolves changing the numerology and maximum transmission bandwidth. IAB-DU uses average RB utilization ratio $\Gamma_{t_1}^m=\frac{1}{t_1}\sum_{t={1}}^{t_1}\Gamma_{t}^n$  and traffic load $\rho_{t_1}^m=\frac{1}{t_1}\sum_{t={1}}^{t_1}\rho_{t}^m$ to calculate the required RB as follows:
\begin{equation}
\beta^{u, \omega}_{t_1}=\lfloor \frac{\Gamma_{t_1}^m+\rho_{t_1}^m}{2}\beta^{u, \omega} \rfloor.
\end{equation} 
 We consider this RB update happens every $t_1$ $ms$.  
\section{Problem Formulation}
\label{sec:problem_formulation} 
We formulate an optimization problem aims to find the optimal set of state-action pairs, i.e., policy $\pi^*$ such that the expected energy consumption is minimized subject to the performance constraints of each microwave node $j$:
\begin{subequations}
\label{eq:problem_formulation4}
\begin{align}
    &\underset{\vect{x} \; \; T\to \infty}{\text{min lim}}\frac{1}{T}\mathbf{E}_{\pi}[\sum_{t^j_m=1}^{T}\sum_{k=0}^{K}\sum_{m=1}^{M^j} x^j_{m, k} \Phi(\Omega^j, \mathcal{A}^j)P^{j}_{m,k} t^j_m]
\tag{\ref{eq:problem_formulation4}}\\
	& \text{subject to: }\nonumber\\
	&\sum_{m=1}^{M^j}x^j_{m, 6} P^{TX}_{m,6} (t^j_m)\leq\sum_{m=1}^{M^j} \tilde{P}^{j}_{m,6}(t^j_m),\label{first:a1}\\
  &\sum_{m=1}^{M^j} x^j_{m,6}D^j_m(t^j_m)\geq D(t^j_m),\label{first:a2}\\
   &\sum_{k=4}^{K}\sum_{m=1}^{M^j} x^j_{m, k} \geq 1. \label{first:a3} 
\end{align}
\end{subequations}

The constraint (\ref{first:a1}) ensures that the transmission power should be less than the total power available in the serving sub-state. The constraint in (\ref{first:a2}) guarantees that the microwave node should satisfy the DL data rate needs to be sent to the IAB network. The constraint in (\ref{first:a3})  ensures that each microwave node should have at least one radio in the serving sub-state of the ON state.

Once the DL data reaches IAB-based FWA network, we formulate the following optimization problem to minimize the energy consumption of each IAB-DU $n$:
\begin{subequations}
\label{eq:problem_formulation5}
\begin{align}
    &\underset{(\vect{y}, \vect{z},\vect{w}, \vect{\kappa}  )}{\min} p^n t
	\tag{\ref{eq:problem_formulation5}}\\
	& \text{subject to: }\nonumber\\
    &p^n \leq p^{Max, n}\label{first:cc1},\\
	&\sum_{v=1}^{|\mathcal{V}^n|}w^{n}_{v}	\beta_{v}^{{n}} (d_{v})  \leq \beta^{u, \omega}_{t_1},\;  \forall n \in \mathcal{N},\label{first:cc2}\\
  &w^{n}_{v} \kappa_{v}^{n}(y^n_{md}+y^n_{mm})\geq1\label{first:cc5}, \\
    &\sum_{v=1}^{V^n}w^{n}_{v}D_v^n\leq D_{Max}^{n, t},\label{first:cc4}\\
  &w^{n}_{v}D_v^n\geq d_v.\label{first:cc3}
\end{align}
\end{subequations}

Based on our system model, the IAB donor is in the middle of  microwave backhaul and IAB-based FWA network. Therefore, we formulated the following objective function and optimization problem that links  microwave backhauk link and the IAB network at the IAB donor. In other words, the following problem combines (\ref{eq:problem_formulation4}) and (\ref{eq:problem_formulation5}) and ensures longhaul microwave backhauling and IAB network satisfy the required data rate.
\begin{subequations}
\label{eq:problem_formulation6}
\begin{align}
    &\underset{(\vect{x},\vect{y}, \vect{z},\vect{w}, \vect{\kappa}))}{\min}  A(\vect{x},\vect{y}, \vect{z},\vect{w}, \vect{\kappa})=\frac{1}{T}\mathbf{E}[\sum_{t^j_m, t=1}^{T}p^n t +  \nonumber\\
    & \sum_{k=0}^{K}\sum_{m=1}^{M^j} x^j_{m, k} \Phi(\Omega^j, \mathcal{A}^j)P^{j}_{m,k} t^j_m]
	\tag{\ref{eq:problem_formulation6}}\\
	& \text{subject to: }\nonumber\\
	&\sum_{m=1}^{M^j} x^j_{m,6}D^j_m(t^j_m)\geq D(t^j_m),\label{first:ccc1}\\
    &\sum_{v=1}^{V^n}w^{n}_{v}D_v^n\leq D_{Max}^{n, t},\label{first:ccc2}\\
  &\sum_{v=1}^{V^n}w^{n}_{v}D_v^n\geq \sum_{v=1}^{V^n}d_v.\label{first:ccc3}
\end{align}
\end{subequations}

These formulated optimization problems can be easily converted to energy efficiency $E_e$ defined in \cite{etsi28.310}, where 
$E_e\text{[bit/Joule]}= \frac{\text{Data rate [bit/s]}} {\text{Energy consumption [Joule/s]}}$. We  will show the energy consumption minimization and its corresponding energy efficiency in performance evaluation.

\section{Proposed Solution}
\label{sec:Proposed_Solution}

To solve the formulated problem in  (\ref{eq:problem_formulation4}), which is a discrete-time Markov decision process, we need to keep tracking $P^{j}_{m,k}$ at each time $ t^j_m$, which is also time and energy consuming. Based on energy consumption during the data gathering discussed in \cite{lindsey2002data}, we define $\xi^j (t^j_m)$ as energy per bit which is measured in terms of joule per bit at microwave node $j$.  Furthermore, we denote $S^j$ as the size of $\Phi(\Omega^j\mathcal{A}^j) $ and $\sum_{m=1}^{M^j} P^{j}_{m,k}$ in terms of bit at microwave node $j$. Then, we define the energy consumption of collecting $\Phi(\Omega^j\mathcal{A}^j) $ and $\sum_{m=1}^{M^j} P^{j}_{m,k}$ as follows:
\begin{equation}
E^j(t^j_m)= S^j\xi^j (t^j_m).
\end{equation}
Then, we update the objective function of (\ref{eq:problem_formulation4}) as follows:
	\begin{equation}
		\setlength{\jot}{10pt}
		\begin{aligned}
			\label{eq:surrogate}
			\tilde{A}^v (\vect{\pi}) = \underset{T\to \infty}{\text{lim}}\frac{1}{T}[\sum_{t^j_m=1}^{T}( E^j(t^j_m)  + \\ \sum_{k=0}^{K}\sum_{m=1}^{M^j} x^j_{m, k} \Phi(\Omega^j, \mathcal{A}^j)P^{j}_{m,k} t^j_m]).
		\end{aligned}
	\end{equation}	
 In other words, instead of minimizing (\ref{eq:problem_formulation4}), we minimize the following problem to find the optimal policy $\vect{\pi}^*$:
	\begin{equation}
		\begin{aligned}
			\label{eq:averageAosurrogate}
			&\vect{\pi}^*=\underset{\vect{\vect{\pi}}}{\text{minimize}}\ \  \tilde{A}^v(\vect{\pi})\\
			&\ \text{subject to: (\ref{first:a1}), (\ref{first:a2}),  and (\ref{first:a3})}.
		\end{aligned}	
\end{equation}
Furthermore, we convert the problem in	(\ref{eq:averageAosurrogate})  to the unconstrained optimization problem by using the Lagrangian method \cite{boyd2011distributed}. We choose the Lagrangian method \cite{boyd2011distributed} over other techniques because it enables us to change (\ref{eq:averageAosurrogate}) to an unconstrained optimization problem by including constraints in the objective function. The unconstrained optimization problem of (\ref{eq:averageAosurrogate}) can be expressed as follows:
	\begin{equation}
		\begin{aligned}
			\mathcal{L(\vect{\pi},\vect{\Lambda}, \vect{\lambda}, \vect{\mu}})= 	\tilde{A}^v(\vect{\pi}) + \\ \vect{\Lambda}(\sum_{m=1}^{M^j}x^j_{m, 6} P^{TX}_{m,6} (t^j_m) - \sum_{m=1}^{M^j} \tilde{P}^{j}_{m,6}(t^j_m))  + \\\vect{\lambda}  (D(t^j_m)- \sum_{m=1}^{M^j} x^j_{m,6}D^j_m(t^j_m)) +  \\ \vect{\mu}(  1-\sum_{k=4}^{K}\sum_{m=1}^{M^j} x^j_{m, k}). 
		\end{aligned}
\label{eq:problem_formulationLagrangianf}
	\end{equation}
We denote $\vect{\Lambda}$  as the Lagrangian multiplier associated with the constraint (\ref{first:a1}), $\vect{\lambda}$ as the Lagrangian multiplier for the constraint in (\ref{first:a2}), and $\vect{\mu}$ as the Lagrangian multiplier associated with the constraint (\ref{first:a3}). Then,  using (\ref{eq:problem_formulationLagrangianf}), we formulate the Lagrange dual function $g(\vect{\Lambda}, \vect{\lambda}, \vect{\mu})$ to find $\vect{\Lambda}$, $\vect{\lambda}$ and $\vect{\mu}$:
	\begin{equation}
		\begin{aligned}
			g(\vect{\Lambda}, \vect{\lambda}, \vect{\mu})= \underset{\vect{\pi}}{\text{inf}}\;	\mathcal{L(\vect{\pi},\vect{\Lambda}, \vect{\lambda}, \vect{\mu}}).
		\end{aligned}
		\label{eq:problem_Lagrangian_duo}
	\end{equation}
	A solution of (\ref{eq:problem_Lagrangian_duo}) should meet the Karush–Kuhn–Tucker (KKT) conditions defined as follows:
	\begin{itemize}
		\item Stationarity: ${ \nabla_{\vect{\pi}} }\mathcal{L(\vect{\pi},\vect{\Lambda}, \vect{\lambda}, \vect{\mu}}) =0.$
		\item Complementary slackness: $\vect{\Lambda}(\sum_{m=1}^{M^j}x^j_{m, 6} P^{TX}_{m,6} (t^j_m) - \sum_{m=1}^{M^j} \tilde{P}^{j}_{m,6}(t^j_m))=0$, $\vect{\lambda} (\sum_{m=1}^{M^j} x^j_{m,6}D^j_m(t^j_m) - D(t^j_m))=0$, and $\vect{\mu} ( \sum_{k=4}^{K}\sum_{m=1}^{M^j} x^j_{m, k} - 1)=0$.
		\item Primal feasibility: $\sum_{m=1}^{M^j}x^j_{m, 6} P^{TX}_{m,6} (t^j_m) \leq \sum_{m=1}^{M^j} \tilde{P}^{j}_{m,6}(t^j_m)$, $\sum_{m=1}^{M^j} x^j_{m,6}D^j_m(t^j_m) \geq D(t^j_m)0$, and $\sum_{k=4}^{K}\sum_{m=1}^{M^j} x^j_{m, k} \geq 1$.
		\item Dual feasibility: $\vect{\Lambda}\succeq0, \vect{\lambda} \succeq0$, and $\vect{\mu}\succeq0$.
	\end{itemize}

Based on KKT conditions, we need to compute optimal $\vect{\Lambda}^*$, 
$\vect{\lambda}^*$, and $\vect{\mu}^*$  for the optimal policy $\vect{\pi}^*$ such that
$\sum_{m=1}^{M^j}x^j_{m, 6} P^{TX}_{m,6} (t^j_m) \leq \sum_{m=1}^{M^j} \tilde{P}^{j}_{m,6}(t^j_m)$, $\sum_{m=1}^{M^j} x^j_{m,6}D^j_m(t^j_m) \geq D(t^j_m))$, and  $\sum_{k=4}^{K}\sum_{m=1}^{M^j} x^j_{m, k} \geq 1$. However, if $\vect{\Lambda}(\sum_{m=1}^{M^j}x^j_{m, 6} P^{TX}_{m,6} (t^j_m) - \sum_{m=1}^{M^j} \tilde{P}^{j}_{m,6}(t^j_m))>0$, $\vect{\lambda} (\sum_{m=1}^{M^j} x^j_{m,6}D^j_m(t^j_m) - D(t^j_m))<0$, and $\vect{\mu} ( \sum_{k=4}^{K}\sum_{m=1}^{M^j} x^j_{m, k} - 1)<0$, there is a violation of KKT conditions. The violation of KKT conditions causes the payment of high penalties $\vect{\Lambda}$,  $\vect{\lambda}$, and $\vect{\mu}$  because of the violation of constraints (\ref{first:a1}), (\ref{first:a2}) ,  and (\ref{first:a3}). Therefore, a feasible policy $\tilde{\vect{\pi}}$  should prevent paying high penalties,  such that: 
\begin{equation}
	\begin{aligned}
			\tilde{A}^v(\vect{\tilde{\vect{\pi}}}) \geq	\mathcal{L(\tilde{\vect{\pi}}, \vect{\Lambda}, \vect{\lambda}, \vect{\mu}})\geq\underset{\vect{\pi}}{\text{inf}}\;	\mathcal{L(\vect{\pi},\vect{\Lambda}, \vect{\lambda}, \vect{\mu}}) =g(\vect{\Lambda}, \vect{\lambda}, \vect{\mu}).
	\end{aligned}
	\label{eq:problem_formulationLagrangian3}
\end{equation}
The condition in (\ref{eq:problem_formulationLagrangian3}) ensures that minimizing  $\vect{\tilde{\vect{\pi}}}$ gives $\vect{\pi}^* \geq g(\vect{\Lambda}, \vect{\lambda}, \vect{\mu})$. Therefore, if $\vect{\pi}^* $ represents an optimal policy, then
\begin{equation}
	g(\vect{\Lambda}, \vect{\lambda}, \vect{\mu}) \leq 	\tilde{A}^v(\vect{\vect{\pi}^*}) \leq 	\tilde{A}^v(\vect{\tilde{\vect{\pi}}}).
\end{equation}
Furthermore, we compute the dual function to obtain $\vect{\Lambda}^*$, $\vect{\lambda}^*$, and $\vect{\mu}^*$:
\begin{subequations}
	\label{eq:problem_formulation1}
	\begin{align}
		&\underset{\vect{\Lambda},\vect{\lambda}, \vect{\mu}}{\text{maximize}}\ \  g(\vect{\Lambda}, \vect{\lambda}, \vect{\mu})
		\tag{\ref{eq:problem_formulation1}}\\
		&\text{subject to:}\nonumber\\
		& \vect{\Lambda}, \vect{\lambda}, \vect{\mu}\succeq 0,\label{first:aa}\\
		& (\vect{\Lambda}, \vect{\lambda}, \vect{\mu}) \in \{\vect{\Lambda}, \vect{\lambda}, \vect{\mu} | g(\vect{\Lambda}, \vect{\lambda}, \vect{\mu}) > - \infty \}. \label{first:bn}
	\end{align}
\end{subequations}
The formulated problem in (\ref{eq:problem_formulation1}) represents the lower bound of  (\ref{eq:averageAosurrogate}). Here, we exclude $-\infty$  to enforce the problem in (\ref{eq:problem_formulation1})  to be feasible.  	Therefore, $\vect{\pi}^*$ represents optimal solution of  $(\ref{eq:averageAosurrogate})$ when $\vect{\Lambda}^*$, $\vect{\lambda}^*$, and $\vect{\mu}^*$  represent the solution of (\ref{eq:problem_formulation1}) such that:
\begin{equation}
	g(\vect{\Lambda}^*, \vect{\lambda}^*, \vect{\mu}^*) \leq 	\tilde{A}^v(\vect{\vect{\pi}^*}).
\end{equation}

To  solve (\ref{eq:problem_formulation5}), our approach starts with minimum numerology and maximum transmission bandwidth configuration available in Table $5.3.2-1$ \cite{etsi138} as initial RBs. Also, IAB station  selects the minimum MCS index and SNR in MCS table \cite{korrai2020ran} to meet the data rate  $d_{v}$ of each terminal $v$. This helps the IAB station to initialize the decision variables $z^ {\omega}_{u}$, $w^{n}_{v}$, $\kappa_{v}^{n}$ and deal with $y^n_{md}$ and $y^n_{mm}$. Furthermore, we can simplify again ($\ref{eq:problem_formulation5}$) by relaxing variables $y^n_{md}$ and $y^n_{mm}$, where $y^n_{md}$ and $y^n_{mm}$ can take values in the closed interval between $0$ and $1$. By using relaxed  $y^n_{md}$ and $y^n_{mm}$, we can use Disciplined Multi-Convex Programming (DMCP) defined in \cite{shen2017disciplined} to solve the problem in ($\ref{eq:problem_formulation5}$). The DMCP principle involves optimizing a specific set of variables in each iteration while keeping the remaining variables fixed. Then, the initial values of variables $y^n_{md}$, $y^n_{mm}$, $z^{\omega}_{u}$, $w^{n}_{v}$,  and $\kappa_{v}^{n}$ are saved in the answer table.

\begin{figure}[t]
	\centering
	\includegraphics[width=1.0\columnwidth]{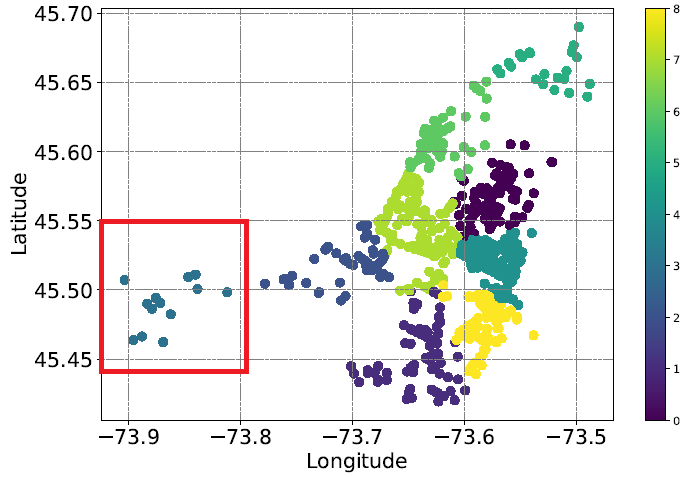}
	\caption{Selected cluster (in red) to represent rural area.}
	\label{fig:Montreal}
\end{figure}
\begin{figure*}[t]
	\centering
	\includegraphics[width=2.0\columnwidth]{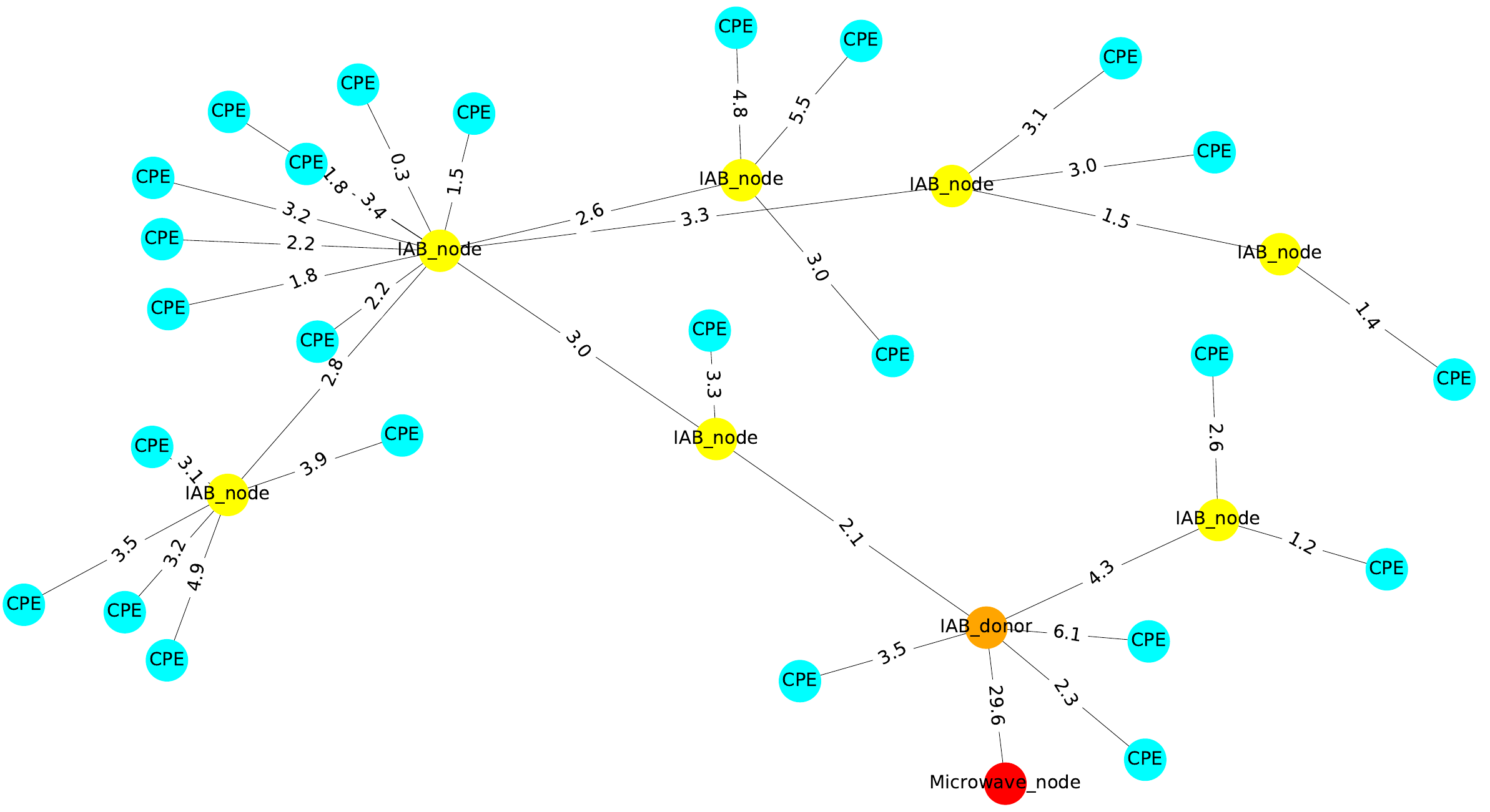}
	\caption{Network topology for rural area and the distance between nodes.}
	\label{fig:IAB_stations}
\end{figure*}
\begin{figure}
	\centering
	\includegraphics[width=1.0\columnwidth]{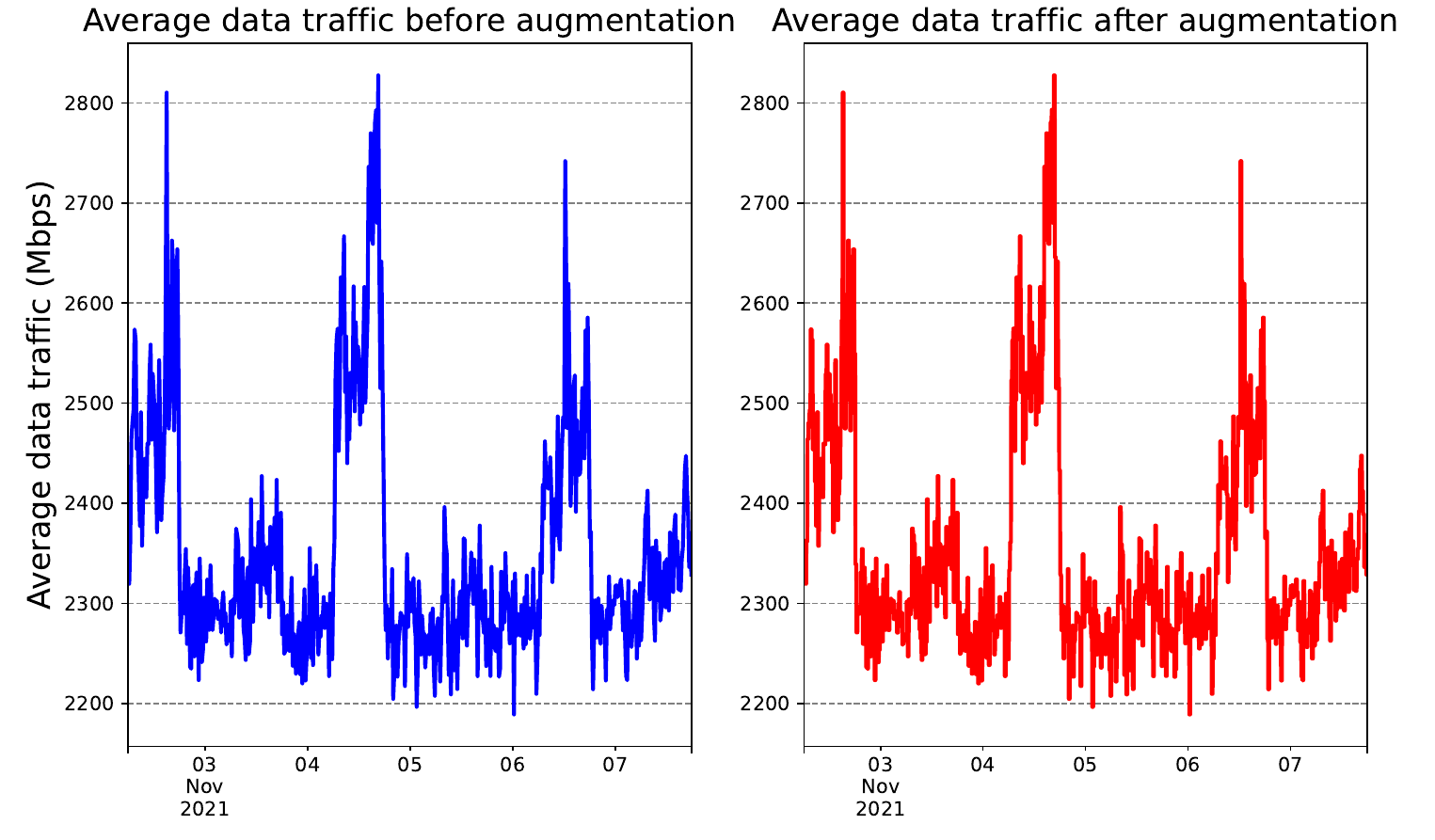}
	\caption{Average backhaul data rate requirement for one week period.}
	\label{fig:IAB_Data_traffic}
\end{figure}
\begin{figure}[t]
	\centering
	\begin{minipage}{0.45\textwidth}
		\centering
		\includegraphics[width=1.0\columnwidth]{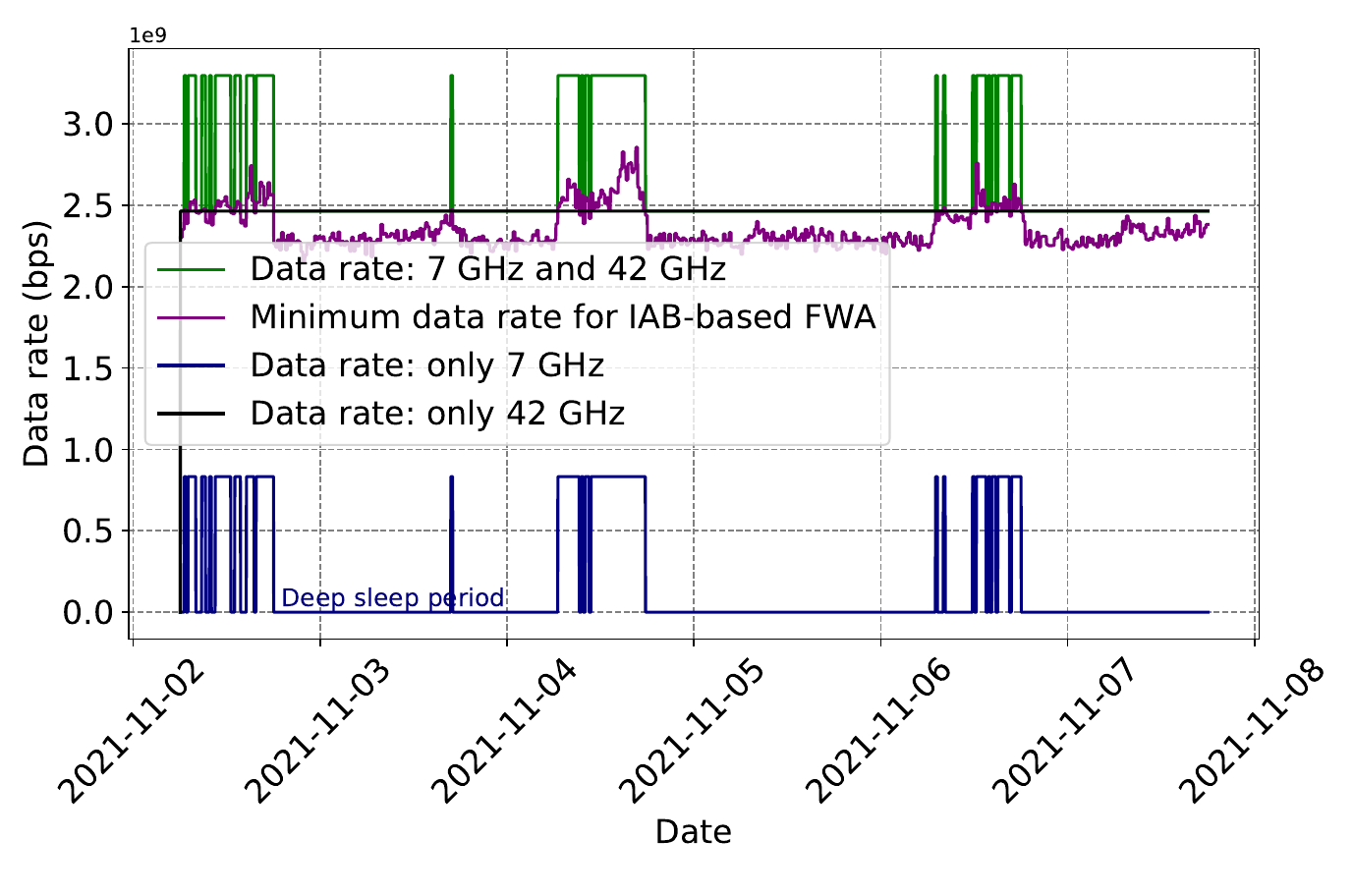}
		\caption{Data rate and microwave radio states.}
		\label{fig:data_rate_backhauling}
	\end{minipage}
	\begin{minipage}{0.45\textwidth}
		\centering
		\includegraphics[width=1.0\columnwidth]{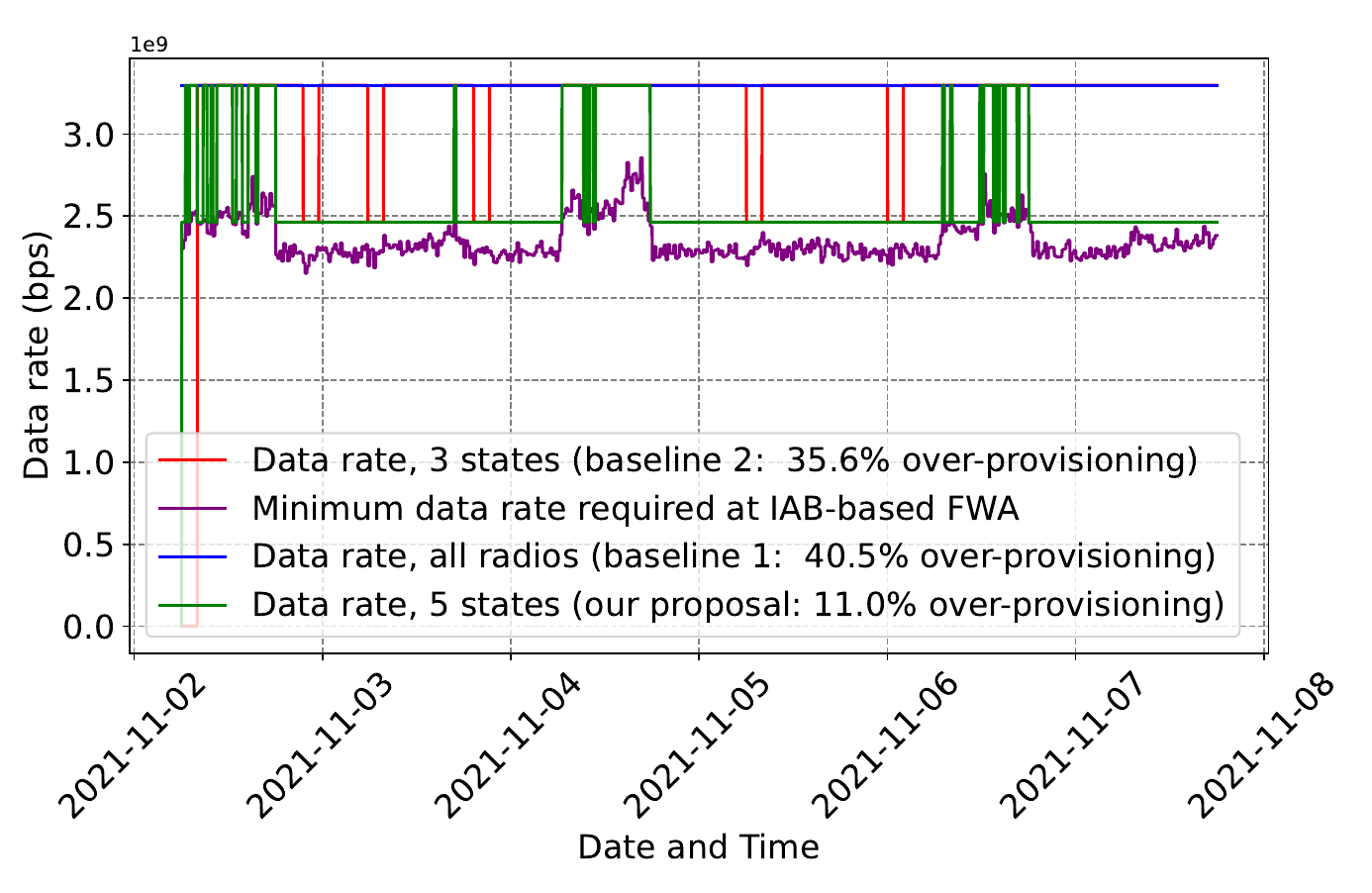}
		\caption{Multi-radio microwave data rates.} 
		\label{fig:Longhaul_microwave_data}
	\end{minipage}
\end{figure}

The IAB station updates $y^n_{md}$, $y^n_{mm}$, $z^{\omega}_{u}$, $w^{n}_{v}$,  and $\kappa_{v}^{n}$  as follows. When $\beta^{u, \omega}_{t_1}- \beta^{u, \omega} >0$, the IAB station changes the values of maximum transmission bandwidth and numerology to increase the number of RB $\beta^{u, \omega}$. However, when $\beta^{u, \omega}_{t_1}- \beta^{u, \omega} <0$, the IAB donor can change the values of maximum transmission bandwidth and numerology to decrease the number of RB $\beta^{u, \omega}$. Otherwise, when  $\beta^{u, \omega}_{t_1}- \beta^{u, \omega} =0$, the IAB station keeps the values of maximum transmission bandwidth and numerology unchanged. In other words,  by keeping $\beta^{u, \omega}$  unchanged.  Furthermore,  the change of MCS index selecting variable is updated based on $\delta^v_{t,f}$, $\delta^{i, n}_{v,t}$,  and $ D_v^n$ as discussed in the previous section. In other words, the IAB station keeps updating the answer table. This process of updating the answer tables corresponds to dynamic programming discussed in \cite{cormen2022introduction}. We choose dynamic programming over other optimization approaches because it can handle inseparable optimization problems, i.e., dependent and overlap optimization problems, where the preliminary result of each formulated problem can be saved in answer tables and then updated to find the solutions.

After solving (\ref{eq:problem_formulation4}) and (\ref{eq:problem_formulation5}), the problem in (\ref{eq:problem_formulation6}) becomes straightforward to solve, as it represents the sum of (\ref{eq:problem_formulation4}) and (\ref{eq:problem_formulation5}).
\begin{figure}[t]
	\centering
	\begin{minipage}{0.45\textwidth}
		\centering
		\includegraphics[width=1.0\columnwidth]{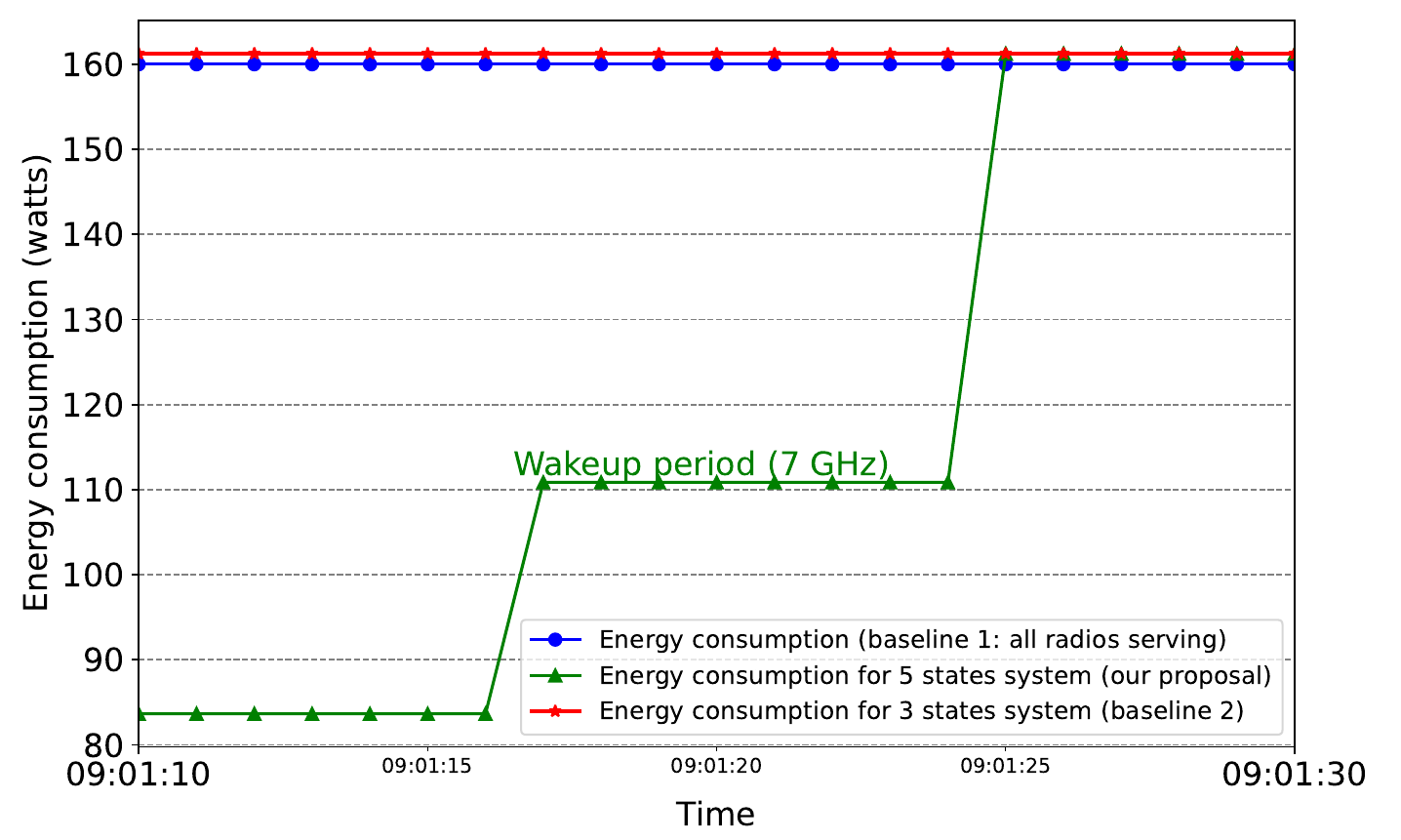}
		\caption{Power consumption during the wake-up.} 
		\label{fig:energy_consuption_second}
	\end{minipage}
	\begin{minipage}{0.45\textwidth}
		\centering
		\includegraphics[width=1.0\columnwidth]{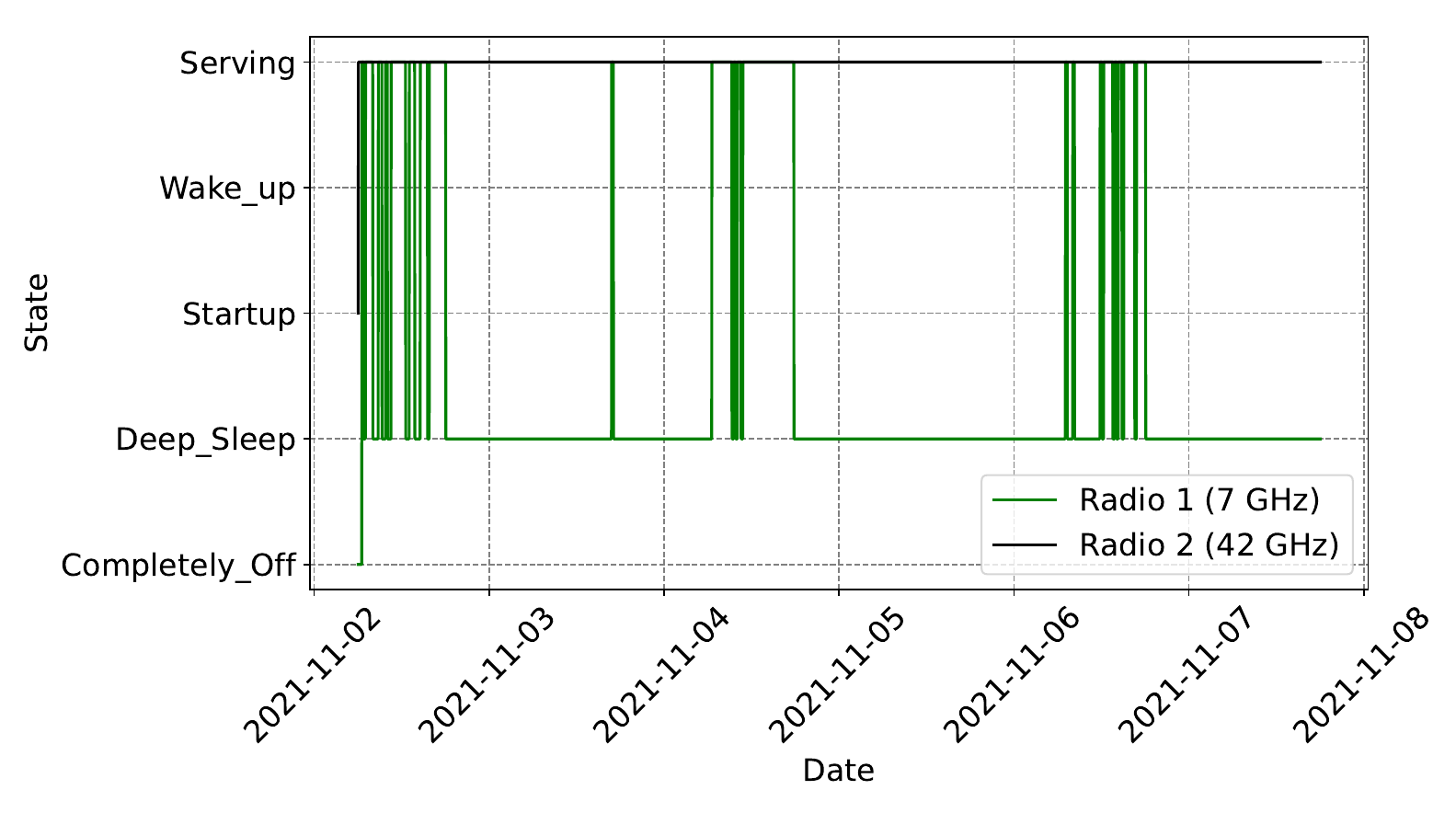}
		\caption{Radio states at microwave node.}
		\label{fig:statesR}
	\end{minipage}
\end{figure}
\begin{figure}[t]
	\centering
	\begin{minipage}{0.45\textwidth}
		\centering
		\includegraphics[width=1.0\columnwidth]{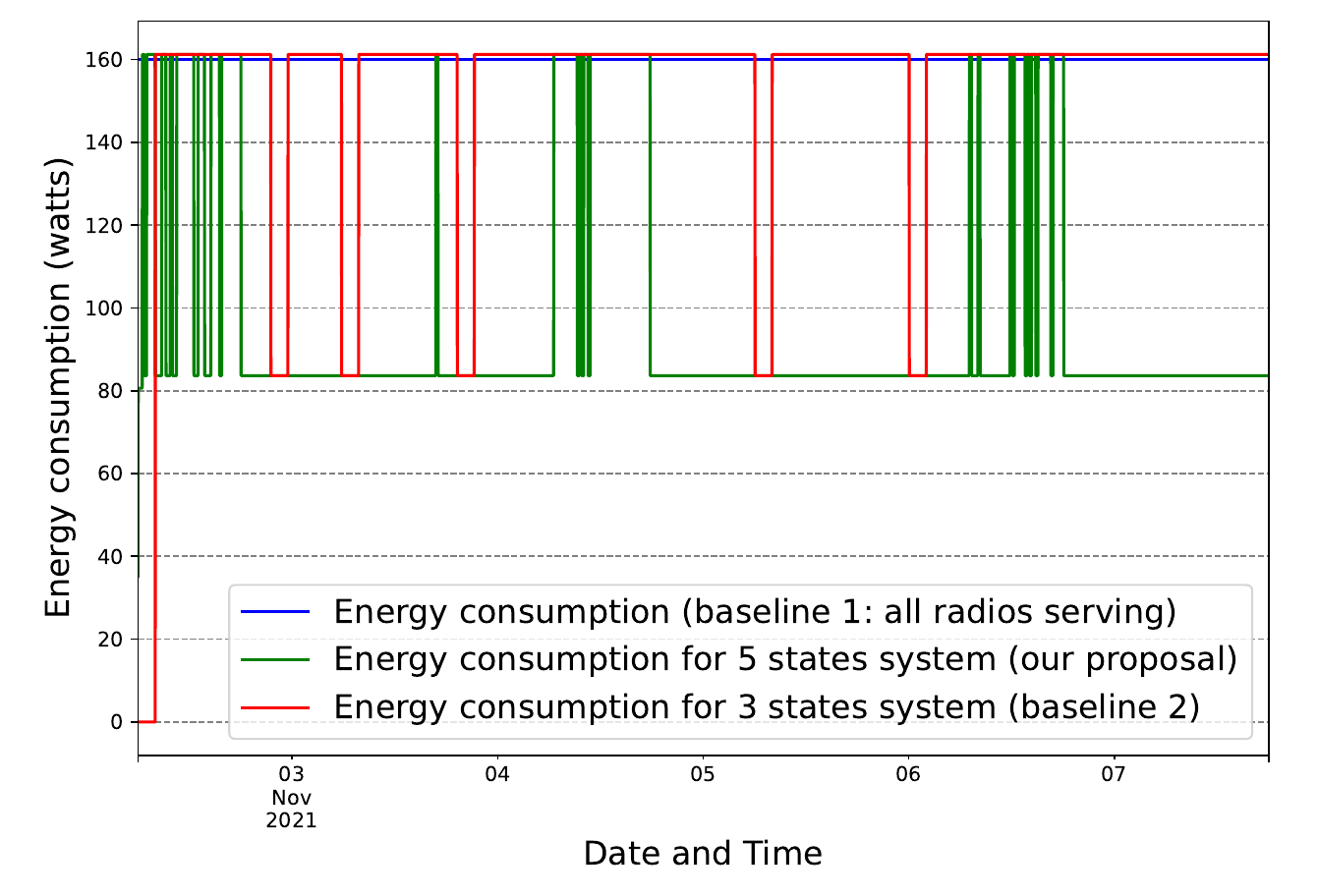}
		\caption{Multi-radio microwave power consumption.}
		\label{fig:energy_consuption_week}
	\end{minipage}
	\begin{minipage}{0.45\textwidth}
		\centering
		\includegraphics[width=1.0\columnwidth]{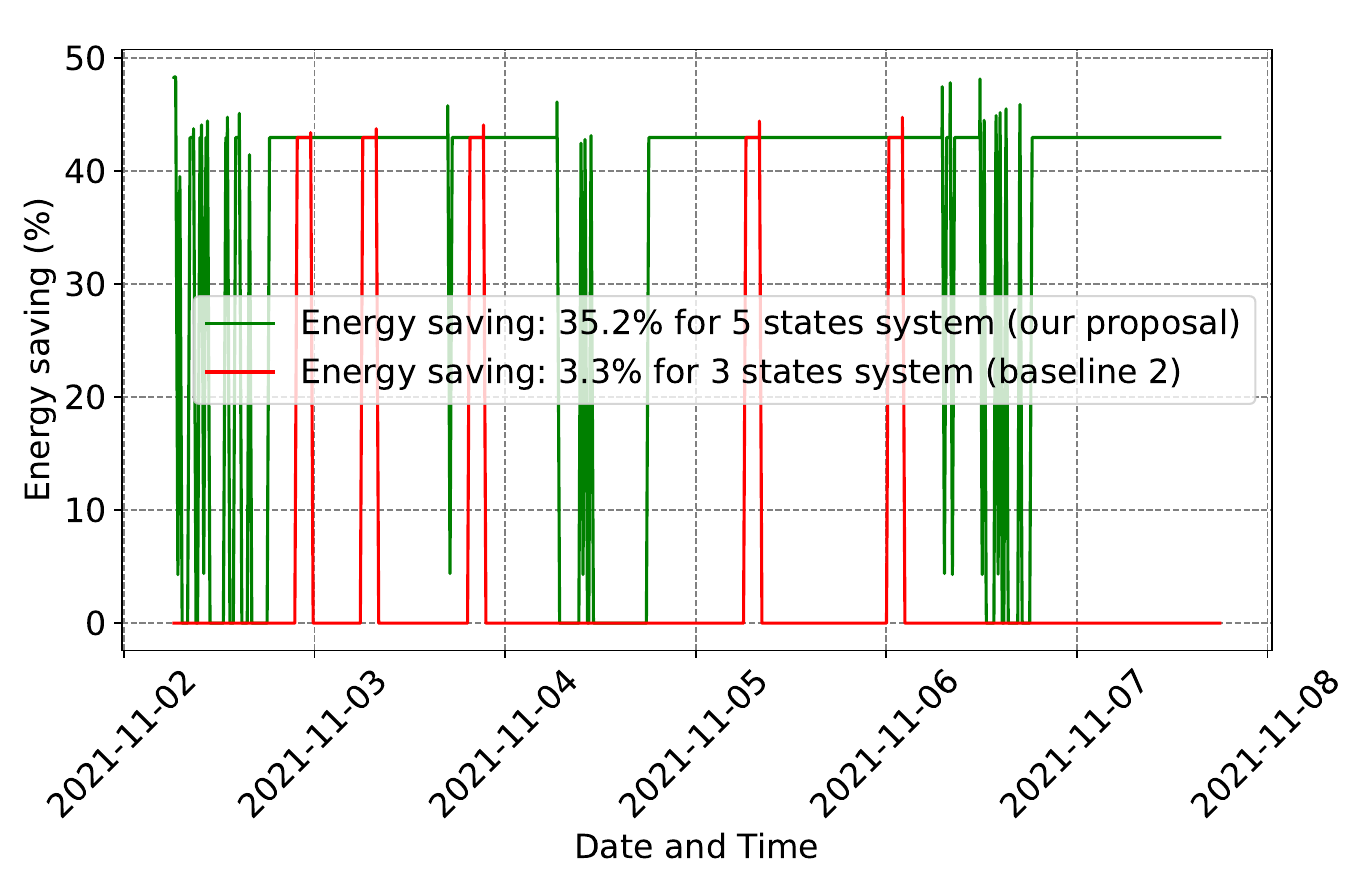}
		\caption{Longhaul microwave energy saving.}
		\label{fig:energy_saving}
	\end{minipage}
\end{figure}
\begin{figure}[t]
	\centering
	\begin{minipage}{0.45\textwidth}
		\centering
		\includegraphics[width=1.0\columnwidth]{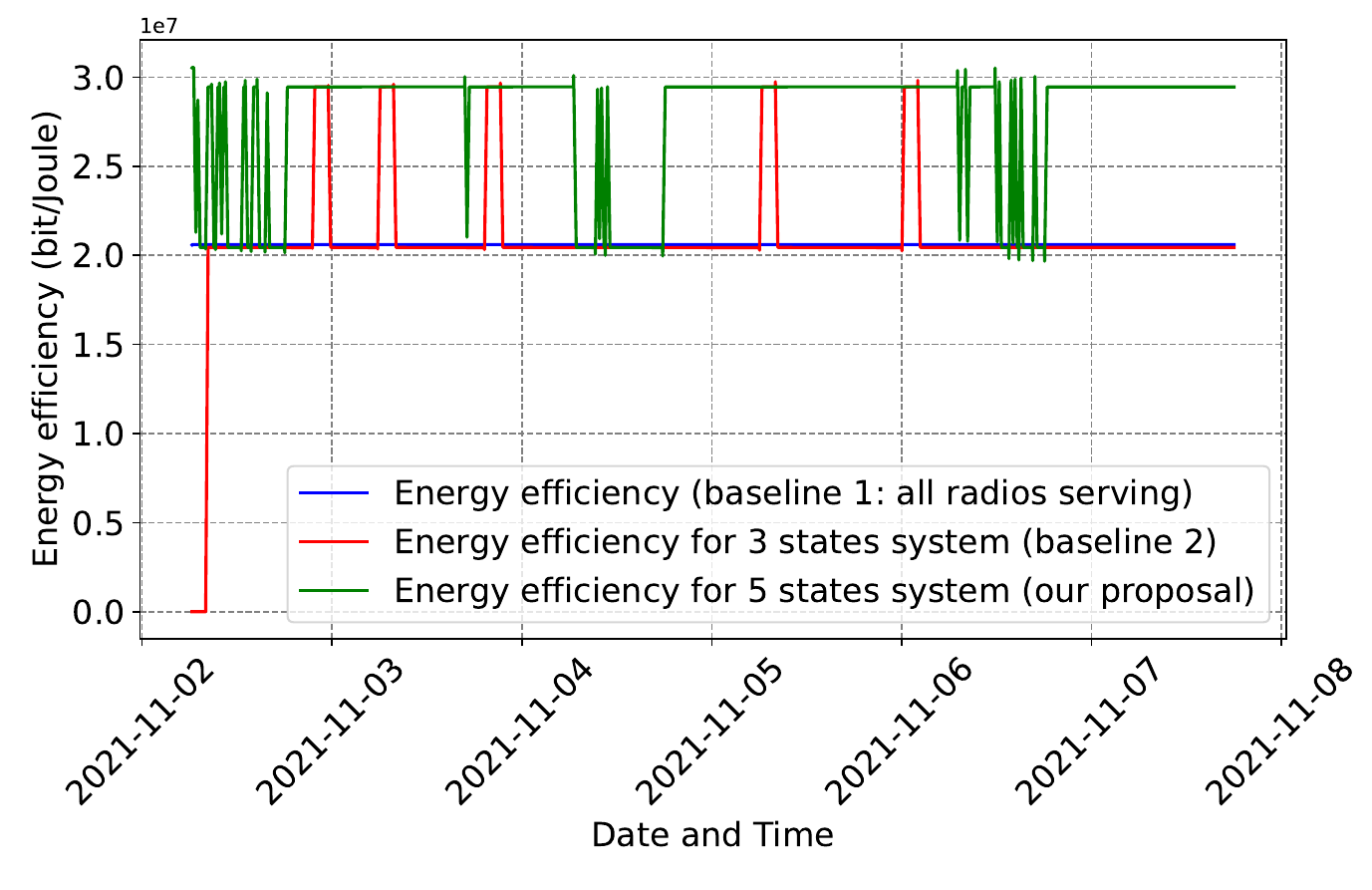}
		\caption{Energy efficiency for microwave backhaul.} 
		\label{fig:energy_efficiency_days}
	\end{minipage}
	\begin{minipage}{0.45\textwidth}
		\centering
		\includegraphics[width=1.0\columnwidth]{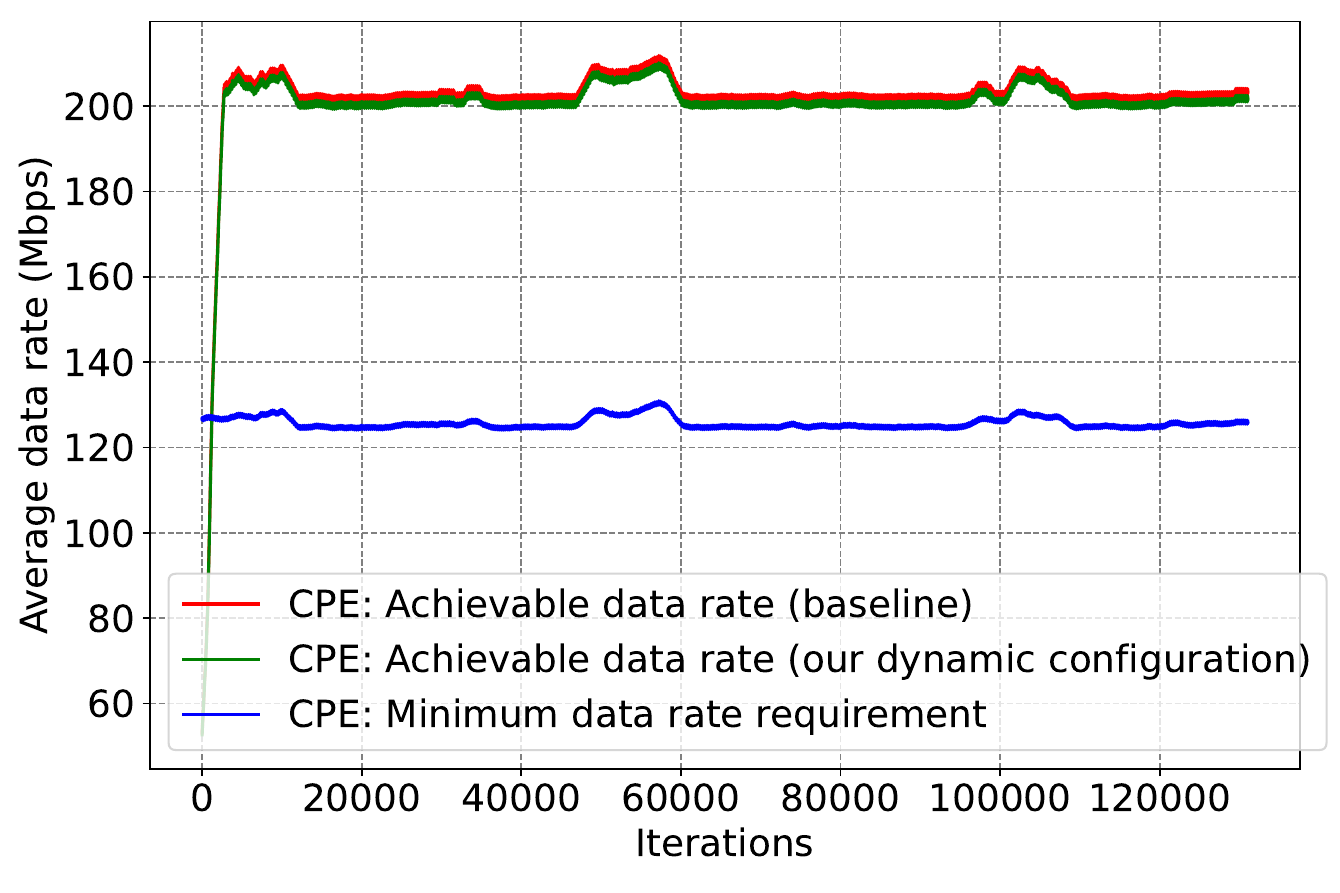}
		\caption{Average achievable data rate per CPE.} 
		\label{fig:DataRateIAB}
	\end{minipage}
\end{figure}	
\begin{figure}[t]
	\centering	
	\begin{minipage}{0.45\textwidth}
		\centering
		\includegraphics[width=1.0\columnwidth]{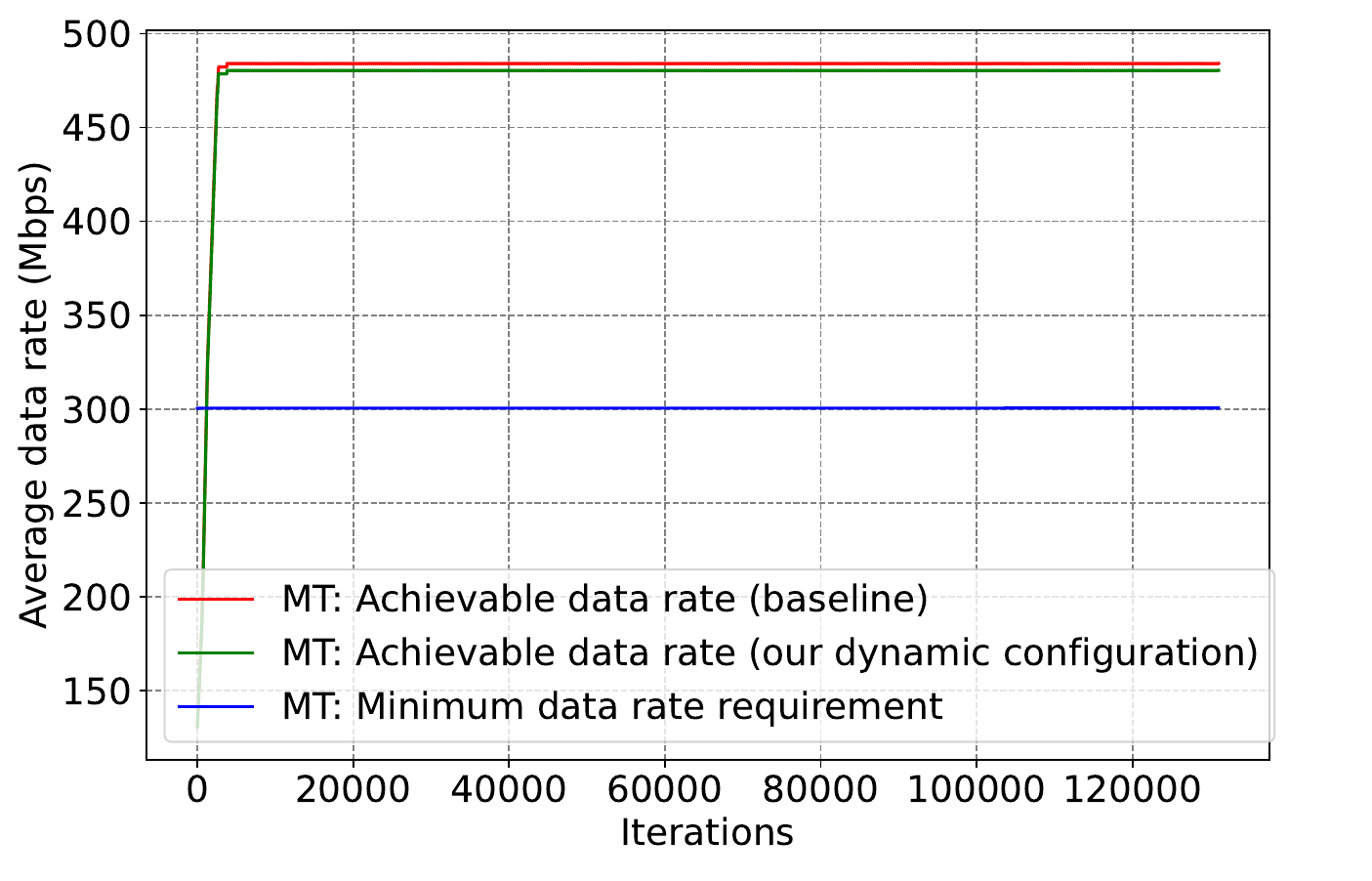}
		\caption{Average achievable data rate per IAB-MT.} 
		\label{fig:DataRateIAB_mt}
	\end{minipage}
\end{figure}
\begin{figure}[t]
	\centering
	\begin{minipage}{0.45\textwidth}
		\centering
		\includegraphics[width=1.0\columnwidth]{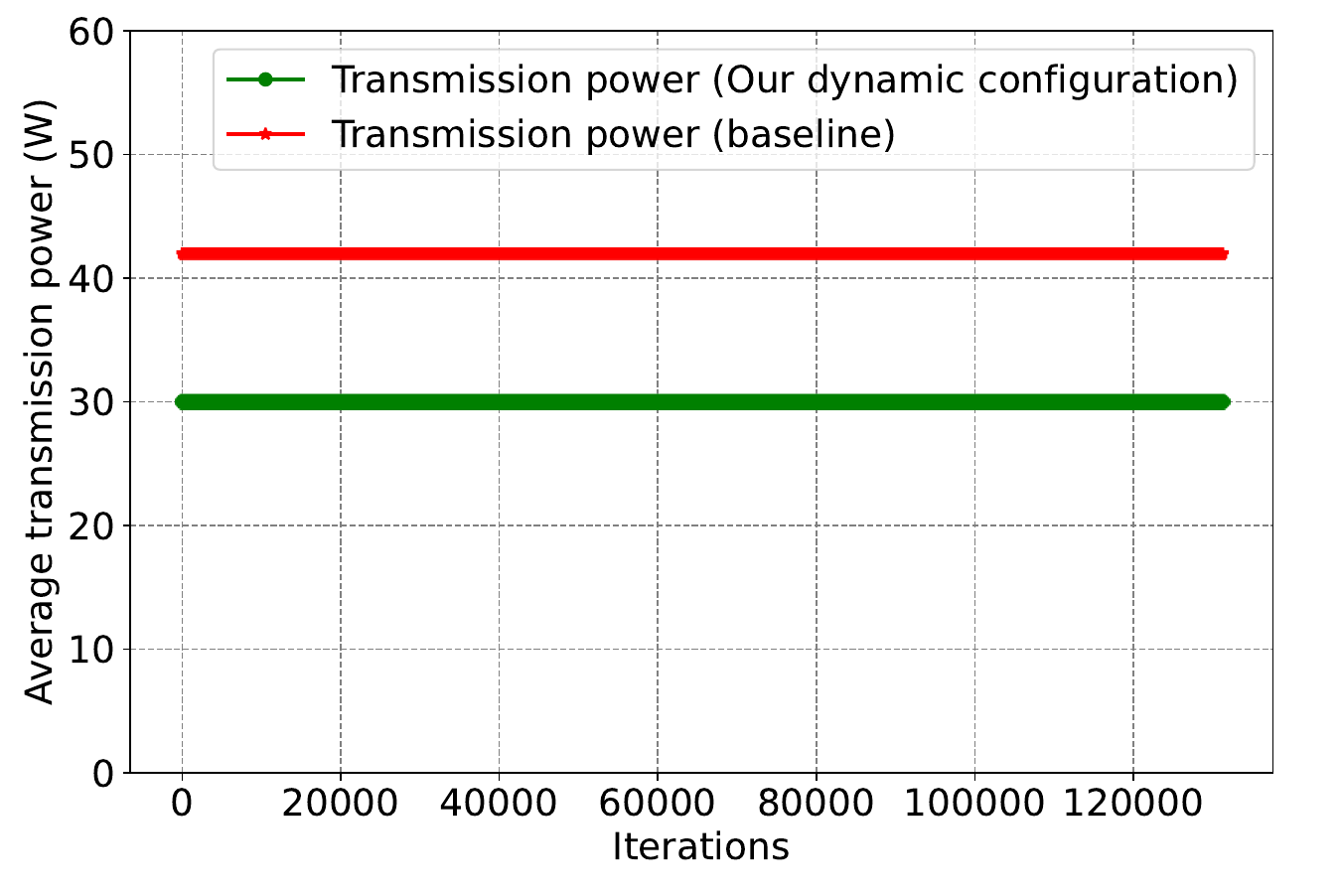}
		\caption{Average transmission power for IAB station.} 
		\label{fig:Transmission_power}
	\end{minipage}
	\begin{minipage}{0.45\textwidth}
		\centering
		\includegraphics[width=1.0\columnwidth]{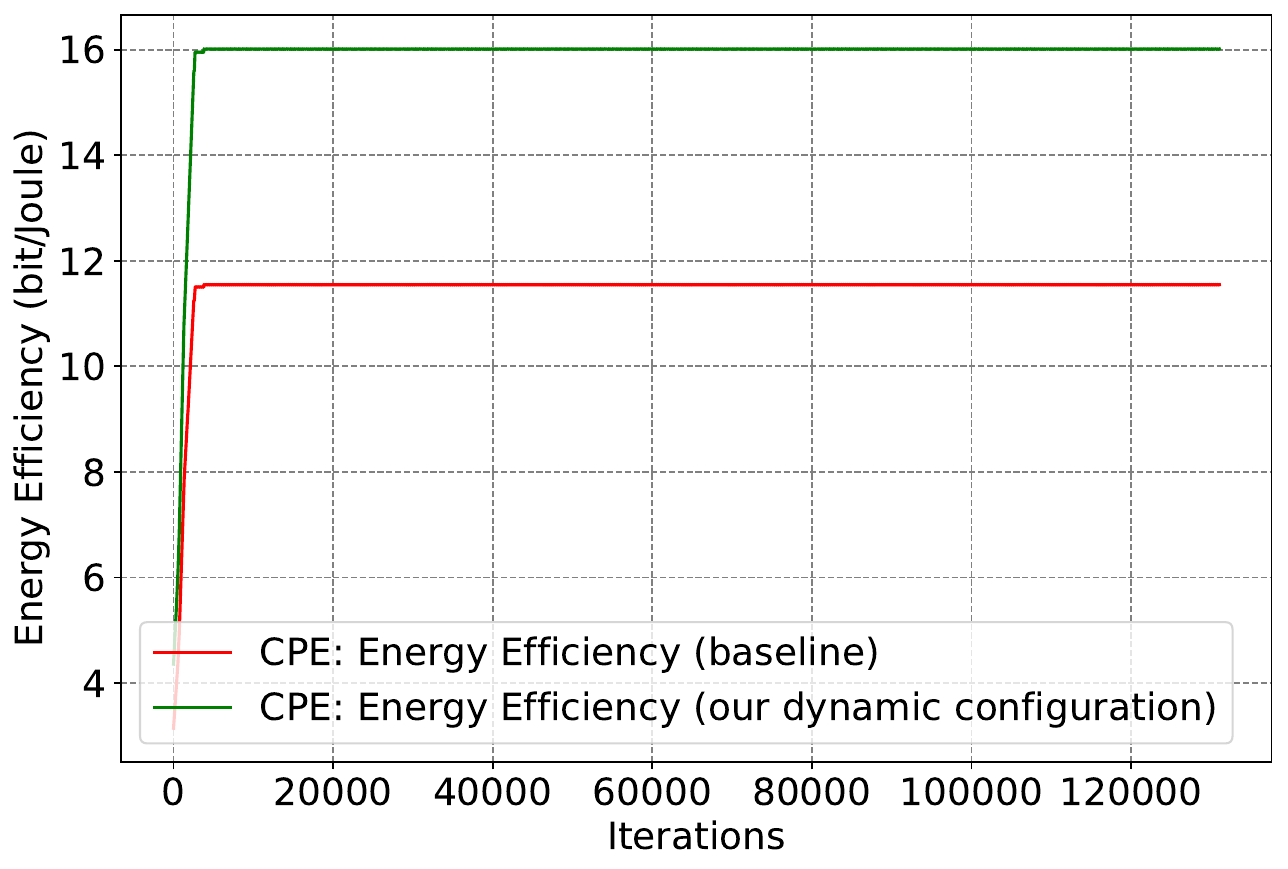}
		\caption{Energy efficiency for IAB-based FWA.} 
		\label{fig:energy_efficiency_IAB}
	\end{minipage}
\end{figure}

\begin{remark}
\emph{Computational complexity of the proposed solution approach:} 
\end{remark}
The overall computational complexity of the proposed solution, which combines dual decomposition and Disciplined Multi-Convex Programming (DMCP) supported by dynamic programming framework, is primarily influenced by the number of optimization variables, Lagrangian multipliers, and the number of iterations.
In problem (\ref{eq:problem_formulation5}), we initialize the decision variables $z^{\omega}_{u}$, $w^{n}_{v}$, and $\kappa_{v}^{n}$, and solve for the binary variables $y^n_{md}$ and $y^n_{mm}$. Once the optimal values of $y^n_{md}$ and $y^n_{mm}$ are determined, we proceed to update the variables $z^{\omega}_{u}$, $w^{n}_{v}$, and $\kappa_{v}^{n}$. Additionally, in problem (\ref{eq:problem_formulation1}), we incorporate the Lagrangian multipliers $\vect{\Lambda}$, $\vect{\lambda}$, and $\vect{\mu}$, along with the variable $x^j_{m,k}$.
Assuming a worst-case scenario with $r$ iterations, the total computational cost can be approximated by $\mathcal{O}(2r^3 + r^2)$,
where $\mathcal{O}(r^3)$ accounts for finding  optimal Lagrangian multipliers $\vect{\Lambda}^*$, $\vect{\lambda}^*$, and $\vect{\mu}^*$. $\mathcal{O}(r^2)$ corresponds to finding the optimal values of $y^n_{md}$ and $y^n_{mm}$. Another $\mathcal{O}(r^3)$ arises from updating decision variables in the answer table. By considering the dominant term, we conclude that the overall computational complexity of our solution approach is
$\mathcal{O}(r^3)$.

The computational complexity of  $\mathcal{O}(r^3)$ means that the running time of our proposed solution grows with the size of the input. This level of complexity is acceptable for rural environments, where network traffic patterns and the operational states of microwave radios vary less frequently than in urban or suburban areas. In contrast, urban and suburban deployments, characterized by higher traffic dynamics and more frequent state transitions, may necessitate alternative approaches, such as fiber-optic-based solutions, rather than  multi-radio microwave backhaul proposed in this work.
 
\section{Performance Evaluation}
\label{sec:simulation_results_analysis}
In this section, we present the simulation setup and performance results of the proposed multi-radio microwave and IAB-based FWA network. Numerical analysis is conducted using Python.
\subsection{Simulation Setup}

To create a multi-radio microwave and IAB-based FWA network topology, we used the transportation dataset from Montreal, Quebec, Canada \cite{motrealData}, covering November 10, 2020, to October 18, 2023, due to the absence of dedicated datasets for this purpose. This dataset records vehicle, cyclist, and pedestrian counts at various intersection points. To simulate a rural environment suitable for multi-radio microwave and IAB-based FWA, we divided the Montreal region into clusters and selected one with sparsely distributed intersections. The selected cluster is shown in Fig. \ref{fig:Montreal}. In the selected cluster,  we retained eight intersection points: one serves as the IAB donor, and the other seven as IAB nodes. Using Sketch, a VeRoViz component described in \cite{veroviz2020}, we defined the cluster boundary.
Inside the selected cluster, we randomly placed 25 nodes representing houses, each equipped with CPE. Assuming signal attenuation with distance, each CPE is connected to its nearest IAB station. A CPE or IAB-MT in close proximity to its parent IAB station utilizes the mmWave band, whereas those located farther away operate in the mid-band. Additionally, we selected an intersection point located far from the cluster boundary to act as a microwave node connected to the IAB donor, forming the microwave backhaul. The resulting network topology and  distances (in kilometers) between nodes  are shown in Fig. \ref{fig:IAB_stations}.

For the multi-radio microwave backhaul and IAB-based FWA setup, we use one microwave node and one IAB donor located 29.6 km apart. Each is equipped with two microwave radios operating at 7 GHz and 42 GHz, with channel bandwidths of 64 MHz and 500 MHz, respectively. Power consumption depends on the radio state: 0 W (completely off), 3 W (deep sleep), 55 W (startup), 50 W (wake-up), and 80 W (serving). But, in practical network, power consumption values are assumed to be obtained via measurement. The IAB donor connects seven IAB nodes to microwave backhaul, which serve 25 houses equipped with CPEs. The distances between CPEs and their respective IAB stations range from 300 meters to 6.1 kilometers. Within the IAB-based FWA network, the frequency bands are 38 GHz for mmWave and 6 GHz for mid-band, with channel bandwidths vary from 10 MHz to 1600 MHz and subcarrier spacings between 15 kHz and 480 kHz \cite{etsi138}. Target SNR values are dynamically selected from Table 2 in \cite{korrai2020ran}. The parameters are set as \(v^{(\zeta)}_{layers} = 4\), \(Q^{(\zeta)}_{mcs} = 8\), \(f^{(\zeta)} = 1.0\), and \(R_{max} = 948/1024\) for calculating $  D_{\text{Max}}^{n, t} $ and $\beta_{v}^{{n}} (d_{v})$.

To model network traffic, we utilize a transportation dataset \cite{motrealData} where  the  number of vehicles, cyclists, and pedestrians is treated
as the number of network packets, each assigned a specific
size. To capture fine-grained traffic dynamics, the original 15-minute resolution dataset is augmented to a 1-second resolution using data augmentation  \cite{maharana2022review}. This augmented dataset reflects the dynamic data rate demands that the multi-radio microwave backhaul must support for the IAB-based FWA network. Fig.~\ref{fig:IAB_Data_traffic} shows the minimum required data rate in Mbps over one week period. Similar analogies between transportation and communication networks have been explored in related works \cite{abdalzaher2019employing, delgado2022network}.

\subsection{ Simulation Results}
Fig. \ref{fig:data_rate_backhauling}  shows the minimum data rate required at the IAB-based FWA.  Using a microwave radio of $42$ GHz, microwave backhaul cannot always satisfy the minimum data rate requirement. We can always meet the minimum data rate by using microwave radios of both $42$ GHz and $7$ GHz. However, when the minimum data rate requirement is low and a microwave radio of $42$ GHz can satisfy it, the microwave radio of $7$ GHz can enter deep sleep state to minimize power consumption and wake up when a higher data rate is needed. When the minimum data rate requirement is high, both microwave radios $42$ GHz and $7$ GHz can be utilized at Microwave node and IAB donor.

We compare our approach with two baselines. The first baseline (Baseline $1$) keeps all microwave radios in the serving state without tracking the power consumed every second. The second baseline, presented in \cite{frithiofson2022energy}, considers $3$ states: deep sleep, completely off, and serving states. In Baseline $2$, when the microwave radio at $7$ GHz reaches $90\%$ of its capacity, the microwave radio at $42$ GHz can enter the serving state to support it, and vice versa. In addition, the microwave radio must remain in any state for at least two hours. Since the Baseline $1$ does not consider the switching of the states, its data rate is almost constant, which is the sum of the data rates of radios that use  $7$ GHz and  $42$ GHz in serving states. In Fig. \ref{fig:Longhaul_microwave_data}, we plot the achievable data rates using our approach and the Baselines $1$ and $2$. This figure shows that the baselines $1$ and $2$ are more over-provisioning compared to our approach, which tries to meet the minimum data rate requirement.

Fig. \ref{fig:energy_consuption_second} shows the power consumption during the wakeup state from deep sleep for the microwave radio of $7$ GHz while  $42$ GHz is serving state. This figure demonstrates that our approach minimizes power consumption when the network is not fully utilized. Furthermore, Fig. \ref{fig:statesR} shows the states of microwave radios in our proposal for a week, while  Fig. \ref{fig:energy_consuption_week} shows the power consumption for one week, indicating that our approach uses less power than the baselines. Also, the results show that the states of microwave radios change less frequently in rural area. However, sometimes our approach and the baseline $2$ consume more power than the baseline $1$ when all microwave radios are in serving states,  due to tracking the power consumption every second.  In other words, the baseline $1$  does not need to track the power consumption every second because it keeps all radios always in the serving states. Fig. \ref{fig:energy_saving} shows that our proposal saves more power than baseline 2 which considers 3 states. In Fig. \ref{fig:energy_saving}, Baseline 1 is not included since it keeps all radios continuously in the serving state, resulting in zero power savings.  Based on the power consumption and data rate, we calculate the energy efficiency in terms of bit/joule for microwave backhaul. The results in Fig. \ref{fig:energy_efficiency_days} show that our proposal is more energy efficient than the baseline.

Figs. \ref{fig:DataRateIAB} and \ref{fig:DataRateIAB_mt} show the average achievable data rate for CPE and IAB-MT in IAB-based FWA connected to microwave backhaul, using mmWave and mid-band with 5G numerologies in our proposed approach. In IAB-based FWA, we compared our proposal against a baseline scenario that uses fixed numerologies and  radio resources (264 RBs for mid-band and 273 RBs for mmWave). Furthermore, Fig. \ref{fig:Transmission_power} illustrates the average transmission power consumed by an IAB station. The results demonstrate that our approach significantly reduces  power consumption, thereby contributing to overall energy savings.
Fig. \ref{fig:energy_efficiency_IAB} presents the energy efficiency of each IAB station, calculated based on power consumption and the achievable data rates for  CPEs . The findings indicate that our proposed method consumes less power than the baseline approach while maintaining a comparable data rate. These results provide a strong proof-of-concept validation for the effectiveness of our method. For future work, we plan to extend the evaluation to larger network scenarios under diverse weather conditions.

\section{Conclusion}
\label{sec:conclution} 
This work presented an energy-efficient multi-radio framework integrating long-haul microwave, IAB, and 5G FWA to enable cost-effective broadband connectivity in rural areas. By addressing the growing energy consumption associated with multi-hop architectures, the proposed solution improves both capacity and energy usage. The framework dynamically adjusts network operation based on traffic demand, optimizing the off, start-up, serving, deep sleep, and wake-up cycles of microwave radios, as well as resource allocation for IAB-based FWA network. The optimization problems were formulated to minimize total energy consumption while satisfying data rate requirements. The formulated optimization problems were efficiently solved using dual decomposition and disciplined multi-convex programming, supported by dynamic programming. Simulation results show that our method significantly reduces power consumption compared to existing solutions while satisfying data rate requirements. Overall, this work offers a sustainable and efficient path for rural broadband deployment using multi-radio microwave and IAB-based FWA  in underserved rural areas.

\bibliographystyle{IEEEtran}

\begin{thebibliography}{10}
	\providecommand{\url}[1]{#1}
	\csname url@samestyle\endcsname
	\providecommand{\newblock}{\relax}
	\providecommand{\bibinfo}[2]{#2}
	\providecommand{\BIBentrySTDinterwordspacing}{\spaceskip=0pt\relax}
	\providecommand{\BIBentryALTinterwordstretchfactor}{4}
	\providecommand{\BIBentryALTinterwordspacing}{\spaceskip=\fontdimen2\font plus
		\BIBentryALTinterwordstretchfactor\fontdimen3\font minus
		\fontdimen4\font\relax}
	\providecommand{\BIBforeignlanguage}[2]{{%
			\expandafter\ifx\csname l@#1\endcsname\relax
			\typeout{** WARNING: IEEEtran.bst: No hyphenation pattern has been}%
			\typeout{** loaded for the language `#1'. Using the pattern for}%
			\typeout{** the default language instead.}%
			\else
			\language=\csname l@#1\endcsname
			\fi
			#2}}
	\providecommand{\BIBdecl}{\relax}
	\BIBdecl
	
	\bibitem{ITU}
	ITU, ``Population of global offline continues steady decline to 2.6 billion
	people in 2023,''
	\url{https://www.itu.int/en/mediacentre/Pages/PR-2023-09-12-universal-and-meaningful-connectivity-by-2030.aspx#:~:text=The%20number%20of%20people%20worldwide,global%20population%20unconnected%20in%202023.},
	2023, published: Sep. 12, 2023, Accessed: Jun. 19, 2025.
	
	\bibitem{Maravedis}
	{Maravedis LLC}, ``{5G} fixed wireless gigabit services today: An industry
	overview,''
	\url{https://shop.maravedis-bwa.com/products/5g-fixed-wireless-gigabit-services-today-an-industry-overview},
	[Online; accessed Sep. 18, 2023].
	
	\bibitem{ericsson2024Mobility}
	{Ericsson}, ``Ericsson mobility report,''
	\url{https://www.ericsson.com/4adb7e/assets/local/reports-papers/mobility-report/documents/2024/ericsson-mobility-report-november-2024.pdf},
	November, 2024.
	
	\bibitem{chaudhuri2021extended}
	K.~R. Chaudhuri, E.~C. Neto, L.~Falconetti, R.~Fassbinder, S.~Guirguis,
	A.~Halder, M.~Irizarry, R.~D. Patel, N.~Saxena, and S.~Sorlescu, ``Extended
	range mmwave for fixed wireless applications,'' in \emph{Proceedings of 97th
		ARFTG Microwave Measurement Conference (ARFTG)}.\hskip 1em plus 0.5em minus
	0.4em\relax IEEE, 2021, pp. 1--4.
	
	\bibitem{hashemi2017integrated}
	M.~Hashemi, M.~Coldrey, M.~Johansson, and S.~Petersson, ``Integrated access and
	backhaul in fixed wireless access systems,'' in \emph{2017 IEEE 86th
		Vehicular Technology Conference (VTC-Fall)}.\hskip 1em plus 0.5em minus
	0.4em\relax IEEE, 2017, pp. 1--5.
	
	\bibitem{3GPP38874}
	{3GPP-TR}, ``Technical specification group radio access network; nr; study on
	integrated access and backhaul (release 16). ({3GPP} {TR} tr 38.874 v16.0.0
	),'' 2018-12.
	
	\bibitem{ericsson2024}
	Ericsson, ``Long haul microwave solutions for high capacities over long
	distances,'' \url{https://www.ericsson.com/en/mobile-transport/long-haul},
	2024, accessed: Jun. 19, 2025.
	
	\bibitem{zu2023arahaul}
	G.~Zu, M.~Nadim, S.~Reddy, T.~U. Islam, S.~Babu, T.~Zhang, D.~Oiao, H.~Zhang,
	and A.~Arora, ``Arahaul: Multi-modal wireless x-haul living lab for
	long-distance, high-capacity communications,'' in \emph{2023 IEEE Future
		Networks World Forum (FNWF)}.\hskip 1em plus 0.5em minus 0.4em\relax IEEE,
	2023, pp. 1--6.
	
	\bibitem{zhang2021resource}
	B.~Zhang, F.~Devoti, I.~Filippini, and D.~De~Donno, ``Resource allocation in
	mmwave {5G} {IAB} networks: A reinforcement learning approach based on column
	generation,'' \emph{Computer Networks}, vol. 196, p. 108248, 2021.
	
	\bibitem{yu2023coordinated}
	M.~Yu, Y.~Pi, A.~Tang, and X.~Wang, ``Coordinated parallel resource allocation
	for integrated access and backhaul networks,'' \emph{Computer Networks}, vol.
	222, p. 109533, 2023.
	
	\bibitem{begishev2021performance}
	V.~Begishev, E.~Sopin, D.~Moltchanov, R.~Pirmagomedov, A.~Samuylov, S.~Andreev,
	Y.~Koucheryavy, and K.~Samouylov, ``Performance analysis of multi-band
	microwave and millimeter-wave operation in 5g nr systems,'' \emph{IEEE
		Transactions on Wireless Communications}, vol.~20, no.~6, pp. 3475--3490,
	2021.
	
	\bibitem{frithiofson2022energy}
	A.~N. Frithiofson, ``Energy efficiency in modern networks,'' \emph{Journal of
		Energy Studies}, vol.~10, no.~2, pp. 123--134, 2022.
	
	\bibitem{ndikumana2024renewable}
	A.~Ndikumana, K.~K. Nguyen, and M.~Cheriet, ``Renewable energy powered and open
	ran-based architecture for 5g fixed wireless access provisioning in rural
	areas,'' \emph{IEEE Transactions on Green Communications and Networking},
	2024.
	
	\bibitem{shen2017disciplined}
	X.~Shen, S.~Diamond, M.~Udell, Y.~Gu, and S.~Boyd, ``Disciplined multi-convex
	programming,'' in \emph{Proceedings of 29th Chinese control and decision
		conference (CCDC)}.\hskip 1em plus 0.5em minus 0.4em\relax IEEE, 2017, pp.
	895--900.
	
	\bibitem{boyd2011distributed}
	S.~Boyd, N.~Parikh, E.~Chu, B.~Peleato, J.~Eckstein \emph{et~al.},
	``Distributed optimization and statistical learning via the alternating
	direction method of multipliers,'' \emph{Foundations and
		Trends{\textregistered} in Machine learning}, vol.~3, no.~1, pp. 1--122,
	2011.
	
	\bibitem{van2012reinforcement}
	M.~Van~Otterlo and M.~Wiering, ``Reinforcement learning and markov decision
	processes,'' in \emph{Reinforcement learning: State-of-the-art}.\hskip 1em
	plus 0.5em minus 0.4em\relax Springer, 2012, pp. 3--42.
	
	\bibitem{rahmawati2022assessing}
	P.~Rahmawati, M.~I. Nashiruddin, A.~T. Hanuranto, and A.~Akhmad, ``Assessing
	3.5 ghz frequency for 5g new radio (nr) implementation in indonesia's urban
	area,'' in \emph{Proceedings of IEEE 12th Annual Computing and Communication
		Workshop and Conference (CCWC)}.\hskip 1em plus 0.5em minus 0.4em\relax IEEE,
	2022, pp. 0876--0882.
	
	\bibitem{lappalainen2022planning}
	A.~Lappalainen, Y.~Zhang, and C.~Rosenberg, ``Planning 5g networks for rural
	fixed wireless access,'' \emph{IEEE Transactions on Network and Service
		Management}, vol.~20, no.~1, pp. 441--455, 2022.
	
	\bibitem{de2022outdoor}
	B.~De~Beelde, Z.~Verboven, E.~Tanghe, D.~Plets, and W.~Joseph, ``Outdoor mmwave
	channel modeling for fixed wireless access at 60 ghz,'' \emph{Radio Science},
	vol.~57, no.~12, pp. 1--14, 2022.
	
	\bibitem{de2023mmwave}
	B.~De~Beelde, M.~Vantorre, G.~Castellanos, M.~Pickavet, and W.~Joseph, ``Mmwave
	physical layer network modeling and planning for fixed wireless access
	applications,'' \emph{Sensors}, vol.~23, no.~4, p. 2280, 2023.
	
	\bibitem{castellanos2023evaluating}
	G.~Castellanos, B.~De~Beelde, D.~Plets, L.~Martens, W.~Joseph, and M.~Deruyck,
	``Evaluating 60 ghz fwa deployments for urban and rural environments in
	belgium,'' \emph{Sensors}, vol.~23, no.~3, p. 1056, 2023.
	
	\bibitem{ndikumana2023digital}
	A.~Ndikumana, K.~K. Nguyen, and M.~Cheriet, ``Digital twin assisted
	closed-loops for energy-efficient open ran-based fixed wireless access
	provisioning in rural areas,'' in \emph{Proceedings of 2023 IEEE Global
		Communications Conference (GLOBECOM)}.\hskip 1em plus 0.5em minus 0.4em\relax
	IEEE, 2023, pp. 6285--6290.
	
	\bibitem{ndikumana2024digital}
	N.~Anselme, K.~K. Nguyen, and M.~Cheriet, ``Digital twin backed closed-loops
	for energy-aware and open ran-based fixed wireless access serving rural
	areas,'' \emph{IEEE Transactions on Mobile Computing}, 2024.
	
	\bibitem{tafintsev2023airborne}
	N.~Tafintsev, D.~Moltchanov, A.~Chiumento, M.~Valkama, and S.~Andreev,
	``Airborne integrated access and backhaul systems: learning-aided modeling
	and optimization,'' \emph{IEEE Transactions on Vehicular Technology}, 2023.
	
	\bibitem{yao2022delay}
	G.~Yao, M.~Hashemi, R.~Singh, and N.~B. Shroff, ``Delay-optimal scheduling for
	integrated mmwave--sub-6 ghz systems with markovian blockage model,''
	\emph{IEEE Transactions on Mobile Computing}, vol.~22, no.~9, pp. 5124--5139,
	2022.
	
	\bibitem{sellin2020enhancing}
	M.~G. Sellin, M.~Edberg, M.~Berggren, M.~Coldrey, J.~Edstam, D.~Eriksson,
	J.~Flodin, J.~Hansryd, A.~Olsson, M.~Ohberg \emph{et~al.}, ``Enhancing 5g
	with microwave,'' \emph{Ericsson, Oct}, 2020.
	
	\bibitem{wang2022triple}
	Y.~Wang, H.~Wu, Y.~Niu, J.~Ding, S.~Mao, B.~Ai, Z.~Zhong, and N.~Wang,
	``Triple-band scheduling with millimeter wave and terahertz bands for
	wireless backhaul,'' \emph{Journal of Communications and Networks}, vol.~24,
	no.~4, pp. 438--450, 2022.
	
	\bibitem{colzani2022long}
	A.~Colzani, M.~Fumagalli, A.~Fonte, A.~Traversa, and E.~Ture, ``Long-reach
	e-band hpa for 5g radio link,'' in \emph{Proceedings of 52nd European
		Microwave Conference (EuMC)}.\hskip 1em plus 0.5em minus 0.4em\relax IEEE,
	2022, pp. 760--763.
	
	\bibitem{etsi138}
	T.~ETSI, ``User equipment ({UE}) radio transmission and reception; part 2:
	Range 2 standalone ({3GPP} {TS} 38.101-2 version 17.6.0 release 17),''
	2022-08.
	
	\bibitem{etsi1308}
	{ETSI-TS}, ``User equipment ({UE}) radio access capabilities ({3GPP} {TS}
	38.306 version 15.2.0 release 15),'' 2018-09.
	
	\bibitem{etsi28.310}
	T.~ETSI, ``{5G}; {LTE};management and orchestration; energy efficiency of {5G}
	({3GPP} {TS} 28.310 version 16.4.0 release 16),'' 2021-04.
	
	\bibitem{lindsey2002data}
	S.~Lindsey, C.~Raghavendra, and K.~M. Sivalingam, ``Data gathering algorithms
	in sensor networks using energy metrics,'' \emph{IEEE Transactions on
		parallel and distributed systems}, vol.~13, no.~9, pp. 924--935, 2002.
	
	\bibitem{korrai2020ran}
	P.~Korrai, E.~Lagunas, S.~K. Sharma, S.~Chatzinotas, A.~Bandi, and
	B.~Ottersten, ``A {RAN} resource slicing mechanism for multiplexing of {eMBB}
	and {URLLC} services in {OFDMA} based {5G} wireless networks,'' \emph{IEEE
		Access}, vol.~8, pp. 45\,674--45\,688, 2020.
	
	\bibitem{cormen2022introduction}
	T.~H. Cormen, C.~E. Leiserson, R.~L. Rivest, and C.~Stein, \emph{Introduction
		to algorithms}.\hskip 1em plus 0.5em minus 0.4em\relax MIT press, 2022.
	
	\bibitem{motrealData}
	{Monteal City}, ``Comptages des véhicules, cyclistes et piétons aux
	intersections munies de feux de circulation,''
	\url{https://donnees.montreal.ca/dataset/comptage-vehicules-pietons},
	October, 2023, [Online; accessed August 27, 2024].
	
	\bibitem{veroviz2020}
	L.~Peng and C.~Murray, ``{VeRoViz}: A vehicle routing visualization toolkit,''
	\url{https://ssrn.com/abstract=3746037}, 2020, accessed: 2021-01-04.
	
	\bibitem{maharana2022review}
	K.~Maharana, S.~Mondal, and B.~Nemade, ``A review: Data pre-processing and data
	augmentation techniques,'' \emph{Global Transitions Proceedings}, vol.~3,
	no.~1, pp. 91--99, 2022.
	
	\bibitem{abdalzaher2019employing}
	M.~S. Abdalzaher and H.~A. Elsayed, ``Employing data communication networks for
	managing safer evacuation during earthquake disaster,'' \emph{Simulation
		Modelling Practice and Theory}, vol.~94, pp. 379--394, 2019.
	
	\bibitem{delgado2022network}
	O.~Delgado, B.~Jaumard, Z.~Ding, F.~Bishay, and V.~Bissonnette, ``A network
	simulator for 5g virtualized networks,'' in \emph{2022 IEEE 8th International
		Conference on Network Softwarization (NetSoft)}.\hskip 1em plus 0.5em minus
	0.4em\relax IEEE, 2022, pp. 237--239.
	
\end{thebibliography}

\end{document}